\documentclass[linenumbers]{aastex631}

\begin{document}
\nolinenumbers
\title{Dynamical Evolution of Four Old Galactic Open Clusters traced by their constituent stars with \textit{Gaia} DR3}

\author[0009-0001-5166-1197]{Shanmugha Balan}
\affiliation{Department of Physics, Birla Institute of Technology and Science, Pilani, 333031, Rajasthan, India}

\author[0000-0001-7470-9192]{Khushboo K. Rao}
\affiliation{Department of Physics, Birla Institute of Technology and Science, Pilani, 333031, Rajasthan, India}
\affiliation{Institute of Astronomy, National Central University, 300 Zhongda Road, Zhongli 32001 Taoyuan, Taiwan}

\author[0000-0003-4662-5463]{Kaushar Vaidya}
\affiliation{Department of Physics, Birla Institute of Technology and Science, Pilani, 333031, Rajasthan, India}

\author[0000-0001-6965-8642]{Manan Agarwal}
\affiliation{Anton Pannekoek Institute for Astronomy \& GRAPPA, University of Amsterdam, Science Park 904, 1098 XH Amsterdam, The Netherlands}

\author[0000-0003-4594-6943]{Souradeep Bhattacharya}
\affiliation{Inter University Centre for Astronomy and Astrophysics, Ganeshkhind, Post Bag 4, Pune 411007, India}



\begin{abstract}
\nolinenumbers

We investigate the evolutionary stages of four open clusters, Berkeley 39, Collinder 261, NGC 6819, and NGC 7789, of ages ranging from 1.6 -- 6 Gyr. These clusters have previously been classified into dynamically young and intermediate age groups based on the segregation level of BSS with respect to red giant branch stars and main sequence stars, respectively. We identify members of these four clusters using the ML-MOC algorithm on Gaia DR3 data. To examine the relative segregation of cluster members of different evolutionary stages, we utilize cumulative radial distributions, proper motion distributions, and spatial distributions in galactocentric coordinates. Our analysis shows that Berkeley 39 and NGC 6819 exhibit moderate signs of population-wise segregation from evolved to less-evolved members. NGC 7789 shows signs of mass segregation only in the cumulative radial distributions. On the other hand, Collinder 261 exhibits high segregation of BSS in the cumulative radial distribution, while other populations show the same level of segregation. 


\end{abstract}




\section{Introduction}
\label{sec:1intro}

In an old stellar system, internal dynamics is the dominant mechanism for mass loss \citep{MeylanHeggie1997}.  A stellar system is led towards energy equipartition through the effect of short- and long-range gravitational interactions called two-body relaxation, eventually resulting in a steady segregation of cluster members based on their masses \citep{Bianchini2016}. At the same time, external factors like the Galactic potential, tidal shocks, and passage through the spiral arms can also play an equally important role in driving the evolution of stellar systems, like open clusters (OCs) \citep{Baumgardt2003}.


These processes significantly affect commonly observed cluster populations, such as the main sequence stars (MSs), main sequence turnoff stars (MSTOs) and red giant branch stars (RGBs), and particulary affect the evolution of binary stars -- leading to the formation of exotic cluster populations \citep{Knigge2009} such as blue straggler stars \citep[BSS;][]{Stryker1993, Bailyn1995}. BSS are typically defined as a brighter and bluer population located along the MS above the MSTO point of a cluster CMD. Based on spectroscopic observations and photometric variability of a few BSS in OCs and globular clusters (GCs), they are found to be as much as twice the mass of turnoff stars in clusters \citep{Shara1997}. According to single stellar isochrones, the masses of BSS are 1 to 2.5 times the mass of MSTOs of their host clusters \citep{Vaidya2022, Panthi2023}. This indicates that BSS are among the most massive populations in a star cluster and are hence subjected to larger gravitational drag \citep{Chandrasekhar1943}, consequently leading to their faster segregation in the cluster center compared to most other populations in a cluster. Using this trait of the BSS, various BSS-based studies have been conducted over the last decade to investigate the dynamical ages of GCs \citep{Ferraro2012, Ferraro2020, Beccari2013, Beccari2023, Sanna2014, Dalessandro2015, Cadelano2022} and OCs \citep{Bhattacharya2019, Rain2020, Vaidya2020, Rao2021, Rao2023, RaoBerkeley172023}.

\cite{Ferraro2012} investigated the dynamical status of GCs using a double normalized BSS radial distribution and sorted them into three families on the basis of their dynamical age -- Family I (dynamically young), Family II (intermediate dynamical age) and Family III (dynamically old). Similar studies have been carried out for OCs using normalized BSS radial distributions using RGBs as a reference population and classified them into different families of dynamical ages \citep{Bhattacharya2019,Rain2020,Vaidya2020}. Later on, \citet{Rao2021} and \citet{Rao2023} studied 23 OCs having a sizeable BSS population ($N_{\text{BSS}} > 10$), by analysing the relative sedimentation level of the BSS using the $A^+_{\text{rh}}$ parameter \citep{Alessandrini2016}, defined as the area between cumulative radial distributions of BSS and a reference population\footnote{\citet{Rao2021} used red giant branch stars (RGBs) and MSTOs as a reference population, and \citet{Rao2023} used MSTOs and MS stars as a reference population.} up to half-mass radius of a cluster. They determined the relationships between $A^+_{\text{rh}}$ and $N_{\text{relax}}$ (cluster age/central relaxation time) as well as the physical parameters for 23 OCs. Their comparison of these correlations in OCs with those in GCs indicated that OCs have a similar relationship to GCs, albeit with higher uncertainties due to the small number of OCs.



Among the studies focusing on OCs, four OCs -- Berkeley 39, Collinder 261, NGC 6819, and NGC 7789 -- exhibit various dynamical stages in comparison to BSS with RGBs or MS stars. These OCs have ages ranging from 1.6 to 6 Gyrs and contain BSS $\ge$ 11. Upon comparing the radial distribution of BSS to RGBs, it has been demonstrated that these OCs are dynamically young \citep{Rain2020,Vaidya2020}. These studies suggest that the most massive population, the BSS, does not show discernible segregation when compared to the most evolved population, the RGBs, possibly due to the small number of BSS in comparison to RGBs. Additionally, these OCs are classified as intermediate age based on the sedimentation level of BSS in comparison to MSTOs and MS \citep{Rao2023}. According to these studies, it is apparent that when comparing BSS with different mass populations, there are varied outcomes regarding dynamical ages of OCs. Additionally, it is also clear that BSS may not be the most optimal approach for determining the dynamical ages of these four OCs. Thus, our goal is to conduct a comprehensive population-wise study in order to explore how different populations are segregated relative to each other. This analysis would allow us understand the mass segregation and dynamical ages of these OCs. We divide cluster members of each evolutionary stages into distinct populations, such as RGBs and SGBs, MSTOs, MS, binaries, and BSS. To carry out this work, we use the spatial and kinematic data of distinct cluster populations from the \textit{Gaia} DR3 \citep{Gaia2023}, the \textit{Gaia}-ESO Survey \citep[GES; ][]{Bragaglia2022} and the WIYN Open Cluster Survey \citep[WOCS; ][]{WOCS2000}.

The rest of the paper is organized as follows: Section \ref{sec:2data} provides the details of the membership determination, differential reddening correction, and the selection of sources for each cluster population in the study. Section \ref{sec:3analysis} describes the methods we use to study mass segregation in these clusters. Section \ref{sec:4discussion} discusses the dynamical state of each cluster based on the properties and observations from the analyses. Finally, Section \ref{sec:5conclusions} concludes the paper.

\section{Data and Membership Identification}
\label{sec:2data}

The \textit{Gaia} mission \citep{Gaia2016} created one of the largest catalogues of astrometric data with precise positions, proper motions, parallaxes, and photometry data of more than one billion stars. \textit{Gaia} has been instrumental in identifying members of OCs \citep{Cantat-Gaudin2022}, their fundamental parameters, and even exotic stellar populations like BSS and yellow stragglers with far more accuracy than ever before \citep{Bhattacharya2019, Rain2021, Vaidya2020}.

\subsection{Membership Determination}

We determine members of the four OCs using the ML-MOC algorithm \citep{Agarwal2021} on \textit{Gaia} DR3 data \citep{Gaia2023}. ML-MOC uses a combination of two unsupervised machine-learning algorithms -- the \textbf{K}-\textbf{N}earest \textbf{N}eighbours algorithm \citep{CoverHart1967} and \textbf{G}aussian \textbf{M}ixture \textbf{M}odels \citep{McLachlanPeel2000} -- on \textit{Gaia} parallaxes and proper motions to determine membership of OCs. The advantage of using ML-MOC is that it uses only proper motions and parallaxes without relying on spatial information and can provide members down to G = 20 mag. Moreover, it does not make any assumptions about the nature or density profile of the cluster. However, ML-MOC sometimes struggles to find all members when the field is heavily crowded. Interestingly, it has been capable of finding many special features of OCs, like tidal tails for many OCs \citep{Bhattacharya2019,Bhattacharya2022} and the bifurcated BSS sequence in Berkeley 17 \citep{RaoBerkeley172023}. \cite{Bhattacharya2022} determined the completeness of ML-MOC for \textit{Gaia} EDR3 data on NGC 2448 by comparing the ML-MOC members with deeper Pan-STARRS1 DR2 imaging data and contamination fraction by comparing ML-MOC members with \textit{Gaia}-ESO Survey data \citep[GES;][]{Bragaglia2022}. Their analysis suggests that ML-MOC is around 90\% complete with a contamination fraction of $\sim2.3$\% down to $G=19.5$ mag. Since parallaxes and proper motions do not change from \textit{Gaia} EDR3 to \textit{Gaia} DR3, members identified using ML-MOC on \textit{Gaia} DR3 and their completeness are as good as \textit{Gaia} EDR3 data. For more details on ML-MOC, readers are referred to \cite{Agarwal2021}. 

To use ML-MOC, generally a search radius of around 1.5 times the tidal radius of the cluster is considered. We download \textit{Gaia} DR3 data for a search radius of $45\arcmin$ for Berkeley 39, $30\arcmin$ for both Collinder 261 \& NGC 6819, and $60\arcmin$ for NGC 7789 from the coordinates of cluster centers. The search radii are chosen while keeping in mind the distance of the cluster and the crowdedness of the field along the direction of the cluster. Using ML-MOC, we identify 867 members in Berkeley 39, 2195 members in Collinder 261, 1365 members in NGC 6819, and 2864 members in NGC 7789. To compare the members obtained with ML-MOC, we use members determined by \cite{VG2023}. They use astrometric parameters -- the proper motion and the parallaxes -- as well as the isochrone shape to identify cluster members. The study reported 923 members in Berkeley 39, 3392 members in Collinder 261, 2535 members in NGC 6819, and 4652 members in NGC 7789, based on sources identified with membership probabilities $> 0.4$. By using a supervised machine learning method, they are able to identify members down to $G=20$ mag, thus obtaining a larger sample with more members. However, this approach did not allow for the identification of exotic populations that deviate from the isochrone, such as the BSS, which are crucial to our study. On the other hand, the ML-MOC algorithm does not rely on isochrones or any other prior information about the cluster. Consequently, it is able to identify exotic stellar populations.

To determine the cluster center for our four OCs (Table \ref{tab:fundamental_params}), we use the Mean Shift algorithm \citep{ComaniciuMeer2002}. The mean shift algorithm is used here because it is a non-parametric algorithm and is well suited for this type of clustered data. Using these cluster centers, we find the radial distances of all the sources from the cluster centers and estimate the radii of the clusters. The distance at which the density of cluster members becomes flat and indiscernible from field stars is considered the cluster radius, $R$, and listed in Table \ref{tab:fundamental_params}.

We utilize radial velocity data from the GES \citep{Bragaglia2022} data for Berkeley 39, from the WOCS \citep{WOCS2000} for NGC 6819 \citep{Milliman2014} and NGC 7789 \citep{Nine2020}, and from \textbf{Gaia} DR3 \citep{Gaia2023} for Collinder 261. We cross-match our members with these catalogues within $2\arcsec$ distance from our members. We threshold the data with $2\sigma$ bounds around the mean radial velocity and remove sources with no information in the dataset. The final fraction of sources found in the catalogues after cross-matching and thresholding is tabulated in Table \ref{tab:galactocentric_coords}. For NGC 6819 and NGC 7789, although we included the binaries as an additional separate population, we could not match enough members as NGC 6819 was observed only down to G = 16.5 mag \citep{Milliman2014} and NGC 7789 down to G = 15 mag \citep{Nine2020}.

\begin{figure}
    \centering
    \includegraphics[width=\linewidth]{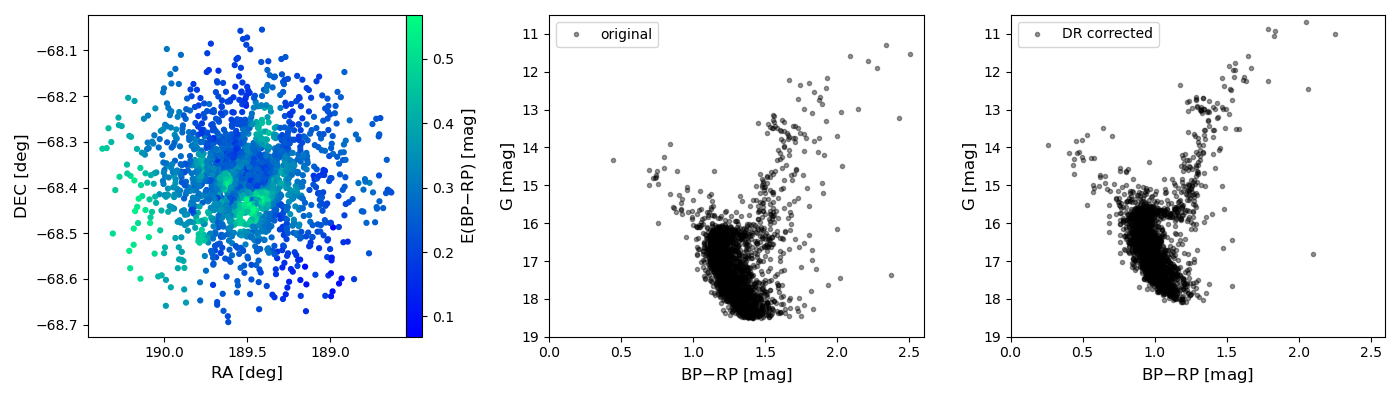}
    \caption{The reddening map (left panel), observed CMD (middle panel), and differential reddening corrected CMD (right panel) for Collinder 261 OC.}
    \label{fig:dr_cr261}
\end{figure}

\subsection{Differential Reddening Correction}
The varying degrees of extinction across a cluster may lead to inaccurate determinations of fundamental parameters and different populations, particularly MS equal mass binaries. To mitigate this issue, we perform differential reddening correction to the identified members following the method from \cite{Rao2023}. First, we select the reddening vector, i.e. R$_{\text{G}}$ = A$_{\text{G}}$/E(BP$-$RP), a quantity corresponding to the direction of the distortion of the red clump on the CMDs. We select R$_{\text{G}}$ = 0.789 from \cite{Rao2023}, which is best fitted to red clump stars across all four OCs. Next, we create a grid over the MS stars with the top border consisting of the reddening vector shifted to the MSTO point of a cluster and the bottom border containing the maximum G magnitude. The right and left borders simply enclose the MS. Using the kNN algorithm, we find 25 spatially closest sources to each cluster member. Of those 25 sources, we compute the mean color $(< BP-RP >)$ and mean magnitude $(<G>)$ of those which fall within the MS grid. The number 25 is selected after a few trials to ensure that each cluster member has a sufficient number of neighboring sources within the main sequence grid. We then calculate the shift along the reddening vector required to match this mean point to the plotted isochrone. With this, we calculate excess magnitude and color, called differential extinction and reddening, respectively. We then subtract differential extinction and reddening from observed $G$ mag and BP$-$RP mag, respectively, to obtain the corrected magnitude and color for all enclosed stars. The reddening maps (left panel), observed CMDs (middle panel), and differential extinction and reddening corrected CMDs (right panel) for Collinder 261 are shown in Figure \ref{fig:dr_cr261} and for the other OCs in Figure \ref{fig:dr_others}. From Figure \ref{fig:dr_cr261} and \ref{fig:dr_others}, we can see a significant reduction in the scatter in the redder part of the observed CMDs and each evolutionary sequence appears sharply defined after the differential reddening correction.

\begin{table*}
	\centering
	\caption{The fundamental parameters of the four OCs: the cluster centers (RA and DEC), ages, metallicities, distances, cluster radii ($R$), core radii ($r_{\text{c}}$), the number of BSS ($N_{\text{BSS}}$), the number of SGB and RGB stars ($N_{\text{SGB\_RGB}}$), the number of MSTO stars ($N_{\text{MSTO}}$) and the total number of member stars identified in this work ($N_{\text{Total}}$).}
	\label{tab:fundamental_params}
	\begin{tabular}{lccccccccccc} 
	\hline
        \hline
                      &          &            &            &                 &        &               &                            &                  &                       &                   & \\
        Cluster       & RA       & DEC        & Age        & [Fe/H] & d     & R        & $r_{\text{c}}$ & $N_{\text{BSS}}$ & $N_{\text{SGB\_RGB}}$ & $N_{\text{MSTO}}$ & $N_{\text{Total}}$ \\	
                      & (deg)    & (deg)      & (Gyr)      & (dex)  & (kpc) & (arcmin) & (arcmin)       &                  & & & \\	
                      &          &            &            &                 &       &          &                &                  &                       &                   & \\
        \hline
                      &          &            &            &                 &       &          &                &                  &                       &                   & \\
        Berkeley 39   & 116.6965 &  $-4.6689$ & 6 & $-0.15$& 4.2   & 14       & 2.2            & 19      & 88           & 149      & 857\\[2pt]
        Collinder 261 & 189.5224 & $-68.3798$ & 6 & $-0.03$& 2.8   & 20       & 2.8            & 36      & 257          & 350      & 2134 \\[2pt]
        NGC 6819      & 295.3285 &  40.1875   &2.7& 0      & 2.5   & 18       & 2.6            & 13      & 59           & 267      & 1350\\[2pt]
        NGC 7789      & 359.3339 &  56.7233   &1.6& 0.04   & 1.9   & 35       & 6.7            & 11      & 159          & 430      & 2838\\[2pt]
                      &          &            &            &                 &       &          &                &                  &                       &                   & \\
        \hline
	\end{tabular}
\end{table*}

\subsection{Selection of Cluster Populations}
\label{sec:selecting-populations}

\begin{figure*}
    \centering
    \includegraphics[width=\linewidth]{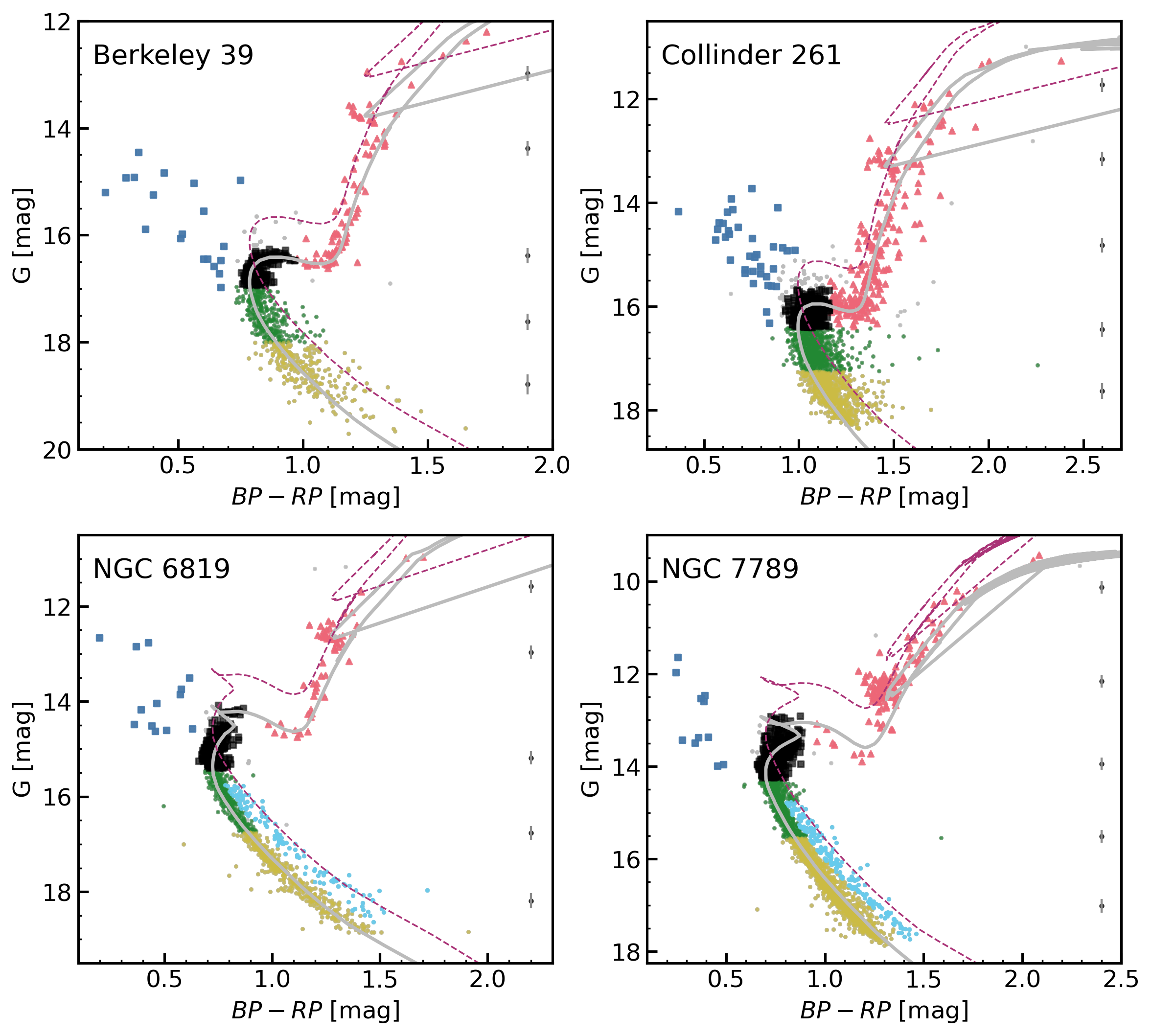}
    \caption{CMDs with fitted single-star PARSEC isochrones (grey solid line) and equal-mass binary isochrones (magneta dashed line). The BSS are shown as blue squares, the RGBs are shown as red triangles, the MSTOs are shown as black squares, and the upper and lower main sequence stars are shown as green and yellow dots. NGC 6819 and NGC 7789 have an equal mass binary track above the MS, and the selected binaries are shown as light blue dots (see Section \ref{sec:selecting-populations} for their selection criterion). Every cluster member is divided into 5 bins based on their magnitude (G) and the average error in G for each bin is calculated. The center of these bins (dark grey dots), and the average error in G scaled $\times 100$ (light grey vertical lines) are shown in the right side of each plot.}
    \label{fig:cmds}
\end{figure*}

To explore the dynamical evolution of the four OCs, we primarily use five different populations in each of the clusters, namely the upper and lower MS, MSTOs, SGBs and RGBs, and BSS. We were able to identify a binary track in the clusters NGC 6819 and NGC 7789, so we added an additional binary population for these two OCs. We identify these populations using the methodology described in \cite{Rao2021}, as follows: we fitted the PARSEC isochrone \citep{PARSEC_Isochrones} for each cluster using distances and metallicties from \cite{Rao2023}, while modified the values of ages and $A_{\text{V}}$ to fit the differential reddening corrected CMDs. The fundamental parameters, such as ages, distances and metallicities of the four OCs are listed in Table \ref{tab:fundamental_params}. The equal mass binary isochrone is plotted by shifting PARSEC isochrones by $G=0.75$ mag and is used to exclude unresolved binaries above the cluster MSTO point, while the ZAMS isochrone is used to remove hot subdwarfs located just below the BSS. To identify the MSTOs, first, the MSTO point of the CMDs is determined, and then the CMDs are normalized by setting the MSTO point at (0, 0). The shifted magnitude and color are represented by $G^*$ and $(Bp-Rp)^*$. Next, a trapezoid is defined with an upper limit at $G = 0.0$ and a lower limit at $G^* = -1.1$ for NGC 6819 and NGC 7789 OCs and $-0.75$ for Berkeley 39 and Collinder 261 OCs. The redward limit is set at $(Bp-Rp)^*=0.2$ and the blueward limit is defined by the equation $G^* = -15\times (Bp-Rp)^* - 1$. Henceforth, we followed the specific steps from \cite{Rao2021} to identify all the populations in our study. We further divided the MS into an upper main sequence (UMS) and lower main sequence (LMS). After some trials, we selected this division by defining a boundary at 1 mag below the MSTO for each cluster. This allows us to obtain a sufficiently large numerical sample which is relatively complete. Figure \ref{fig:cmds} shows the CMDs of the four OCs with plotted isochrones, equal mass isochrones, and the identified populations. To identify binaries in NGC 6819 and NGC 7789, we plotted isochrones corresponding to various mass ratios (q)\footnote{q refers to the ratio of masses between the two stars in an unresolved binary, so an equal mass binary isochrone would correspond to $q=1$.} over the CMDs \citep{LiBinaries2020} and find that the visible binary sequence could be well separated when we choose $q \geq 0.8$ for NGC 6819 and $q \geq 0.9$ for NGC 7789.

Therefore, we select binaries as sources redder than the $q = 0.8$ (or $q = 0.9$ isochrone for NGC 7789) and fainter than $G \geq 15.65$ mag for NGC 6819 (or $G \geq 14.75$ mag for NGC 7789) down till the faintest magnitude limit of the cluster members ($G \leq 19$ mag for NGC 6819 and $G \leq 18$ mag for NGC 7789). We identify 119 binaries in NGC 6819 and 214 in NGC 7789.

\section{Investigating mass segregation}
\label{sec:3analysis}

\subsection{Cumulative Radial Distribution}

\begin{figure*}
    \centering
    \includegraphics[width=\linewidth]{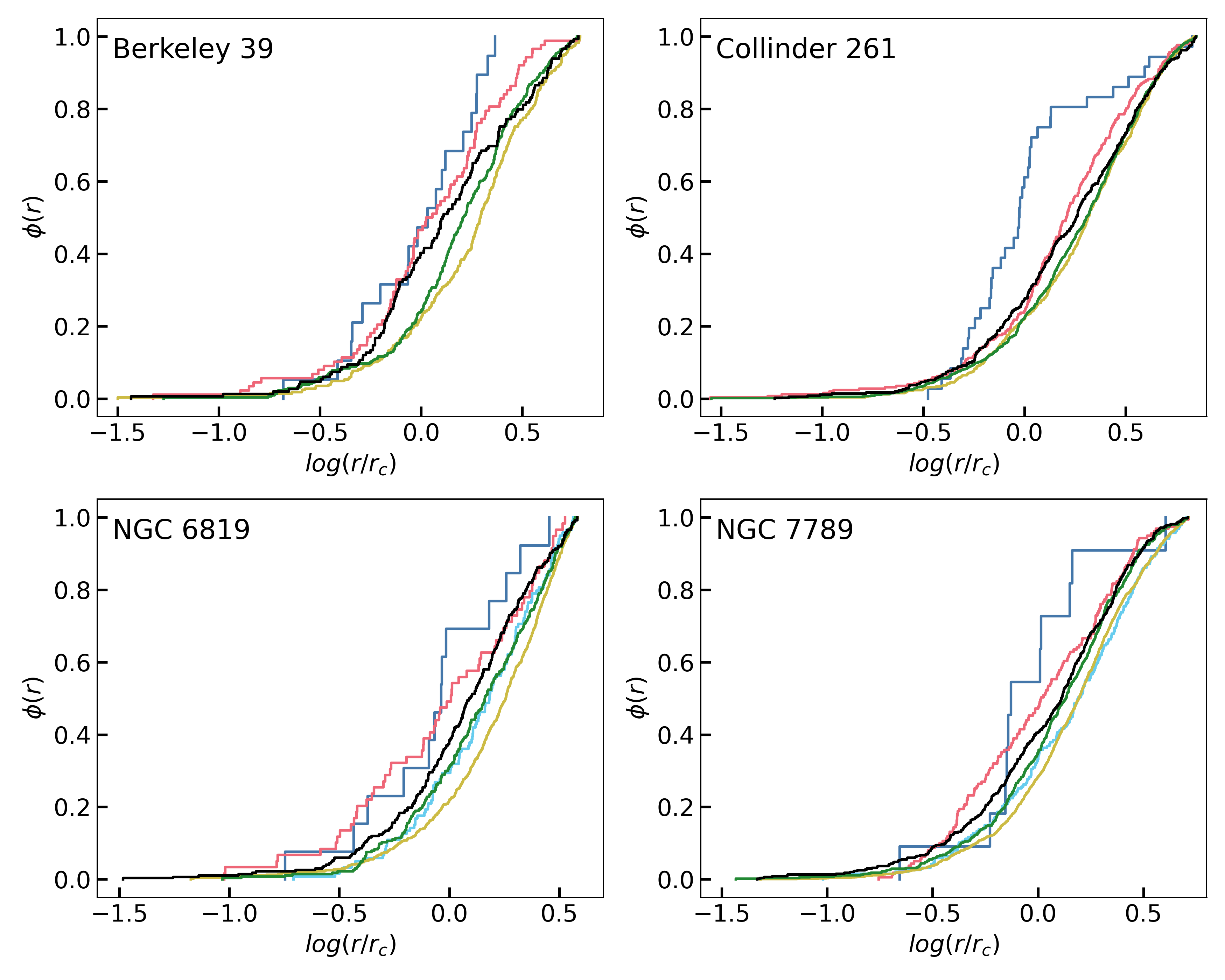}
    \caption{CRDs of the four clusters plotted as a fraction of the sources in each population against the distance from the cluster center normalized by the core radius. The BSS are shown in deep blue, RGBs in red, MSTOs in black, UMS in green, and LMS in yellow. For NGC 6819 and NGC 7789, the additional binary populations are shown in light blue.}
    \label{fig:crds}
\end{figure*}

In order to check the spatial distributions of each cluster population identified in Section \ref{sec:selecting-populations}, we plot their cumulative radial distributions (CRDs). We use the BSS, RGBs, MSTOs, UMS, LMS and binaries for NGC 6819 and NGC 7789 for the four clusters, as shown in Figure \ref{fig:crds}. The radial distances of sources are expressed in the units of core radii $r_{\text{c}}$ of their host clusters, adopted from \cite{Rao2023} and listed in Table \ref{tab:fundamental_params}.



In order to check if the different populations in our study follow different distributions, we compare the CRDs of two different populations pair-wise using the two-sample Anderson-Darling test (henceforth, AD test, first described in \citealt{ADTest}, also see Section 3.4 in \citealt{Bhattacharya2021ADTest} for more details). The AD test highlights the uniqueness of a cluster population in comparison with another. A significance level is calculated by comparing the test statistic against a critical value. To reject the null hypothesis (i.e., both samples are drawn from different distributions), the computed significance level must be under 5\%. However, for significance levels over 5\%, no conclusions can be drawn. The AD test results are listed in Table \ref{tab:ad_test} and discussed in detail in Section \ref{sec:4discussion}. 


\begin{table}
\centering
\caption{The significance levels (in \%) from the AD test for pairwise comparison for every population in each cluster are listed here. We consider the pair of CRDs compared to come from different populations if the significance level is less than 5\% (denoted by italics). The significance values are floored at 0.1\% and capped at 25\%.}
\label{tab:ad_test}

\begin{tabular}{ccccc}
\hline
\hline
Populations  & Berkeley 39 & Collinder 261 & NGC 6819 & NGC 7789 \vrule depth 2ex height 4ex width 0pt\\
\hline
\vrule height 4ex width 0pt
BSS -- RGB  &            {25} &  \textit{{0.1}} &            {25} & {25}            \\
BSS -- MSTO &         {12.64} &  \textit{{0.1}} &            {25} & {20.12}         \\
BSS -- UMS  & \textit{{0.63}} &  \textit{{0.1}} & {\textit{4.28}} & {9.81}          \\
BSS -- LMS  & \textit{{0.10}} &  \textit{{0.1}} & \textit{{0.48}} & \textit{{1.91}} \\
            &                        &                        &                        &                        \\
RGB -- MSTO &         {14.91} &         {18.26} &          {9.18} & {10.35}         \\
RGB -- UMS  & \textit{{0.1}}  & \textit{{0.13}} & \textit{{0.25}} & \textit{{0.2}}  \\
RGB -- LMS  & \textit{{0.1}}  &  \textit{{0.1}} &  \textit{{0.1}} & \textit{{0.1}}  \\
            &                        &                        &                        &                        \\
MSTO -- UMS & \textit{{0.64}} & {\textit{3.12}} & \textit{{2.33}} & {8.6} \\
MSTO -- LMS & \textit{{0.1}}  & \textit{{1.76}} &  \textit{{0.1}} & \textit{{0.1}}  \\
            &                        &                        &                        &                        \\
LMS -- UMS  & \textit{{4.80}} &            {25} & \textit{{0.27}} & \textit{{0.1}} \\
            &                        &                        &                        &                        \\
{BIN -- BSS}  & {--}   & {--}            & {\textit{4.82}} & {5.70}          \\
{BIN -- RGB}  & {--}   & {--}            & {\textit{0.65}} & {\textit{0.1}}  \\
{BIN -- MSTO} & {--}   & {--}            &          {6.93} & {\textit{0.4}}  \\
{BIN -- UMS}  & {--}   & {--}            &            {25} & {4.84}         \\
{BIN -- LMS}  & {--}   & {--}            & {\textit{4.09}} & {25}            \\[4pt]
\hline
\end{tabular}
\end{table}

\subsection{Proper Motion Distribution}

\begin{figure*}
    \centering
    \includegraphics[width=\linewidth]{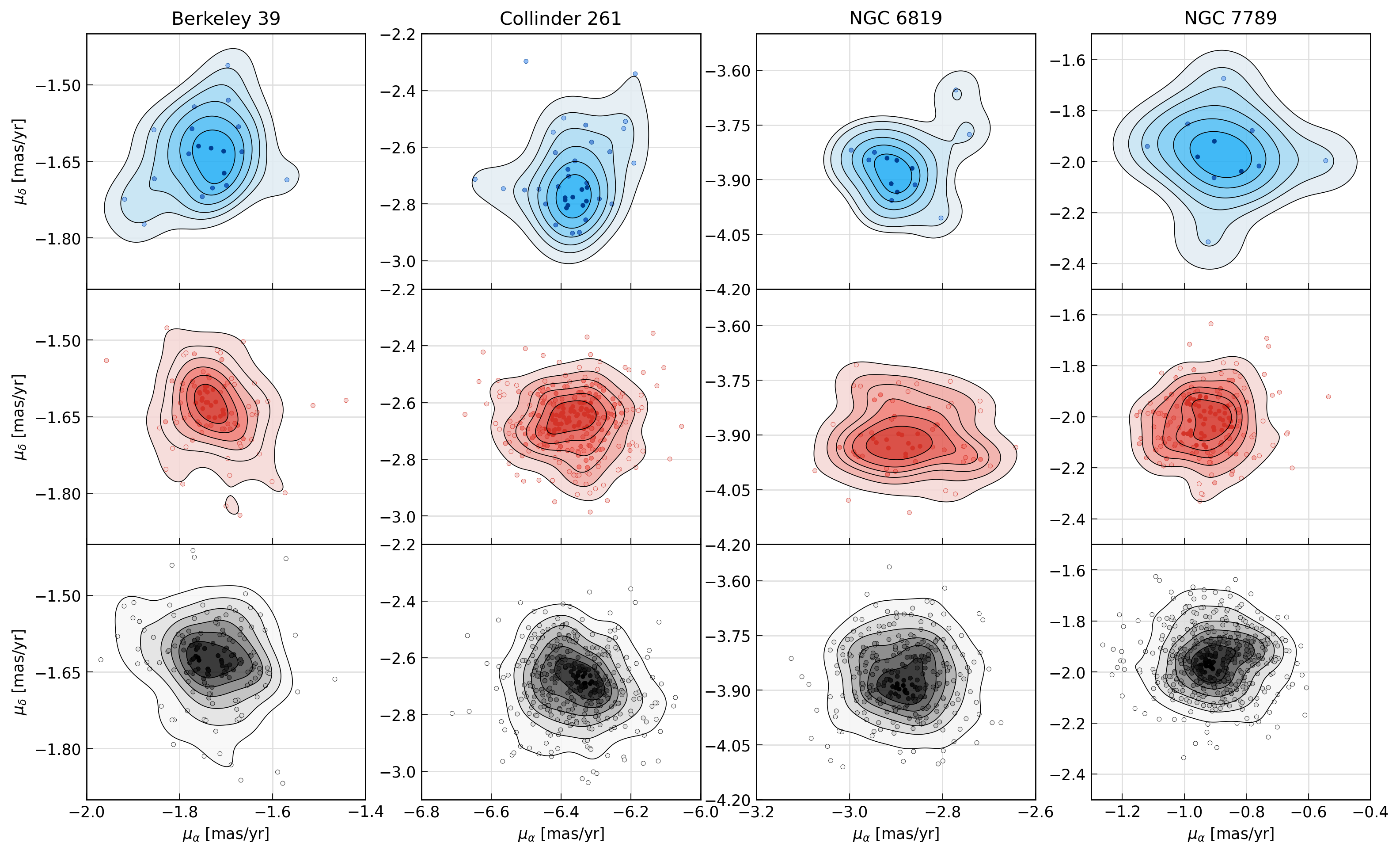}
    \caption{Proper motion distribution with overlaid density contour plots of the four clusters. The BSS are shown in blue, RGBs in red and MSTOs in black. The populations are arranged in decreasing order of mass from top to bottom.}
    \label{fig:pm_scatter}
\end{figure*}

Having compared the cluster populations in position space, the next step is to compare them in the proper motion space. For this, we plot proper motion scatter diagrams for only three populations -- BSS, RGBs, and MSTOs. However, we do not include LMS and UMS stars for this analysis, as errors in their proper motions are almost of the order of the range of proper motions observed in their host clusters.

To visually distinguish the level of segregation better, we use the 2D Gaussian kernel density estimator from the Python package Scipy \citep{SciPy}. The density and concentration of sources towards the mean proper motion are more evident upon overplotting contours in the scatter plot. We plot six contours corresponding to seven equally spaced density levels and the resulting illustration is shown in Figure \ref{fig:pm_scatter}.

\subsection{Spatial and Kinematic distribution in Galactocentric coordinates}

\begin{figure*}
    \centering
    \includegraphics[width=\linewidth]{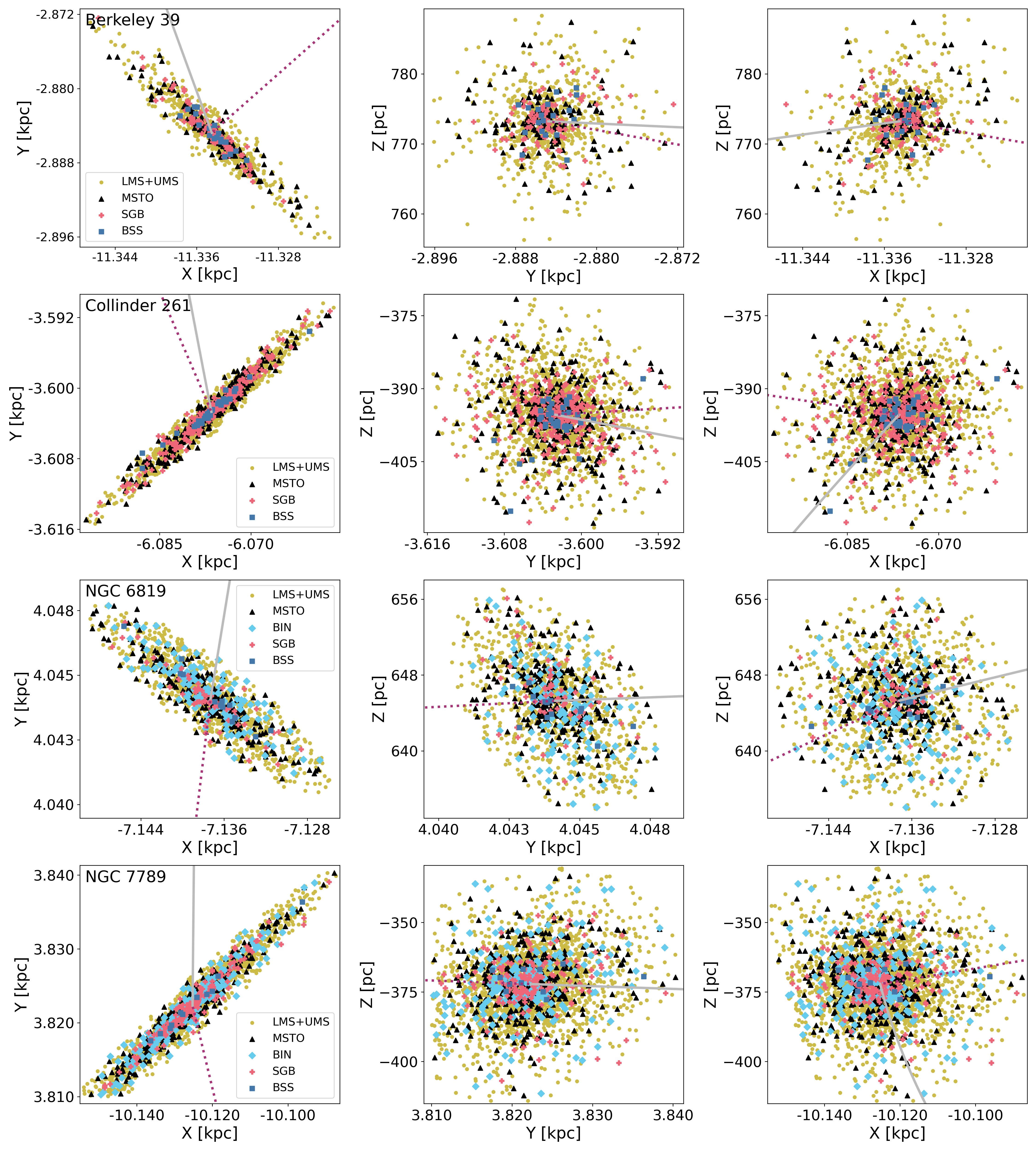}
    \caption{Positions of sources in Berkeley 39, Collinder 261, NGC 6819 and NGC 7789 in galactocentric coordinates, \textbf{arranged row-wise}. Different populations are represented by colored dots as per the inset legend, while the black dotted line represents the line-of-sight to the cluster and the solid line represents the cluster orbital direction.}
    \label{fig:galactocentric-positions}
\end{figure*}

\begin{figure*}
    \centering
    \includegraphics[width=\linewidth]{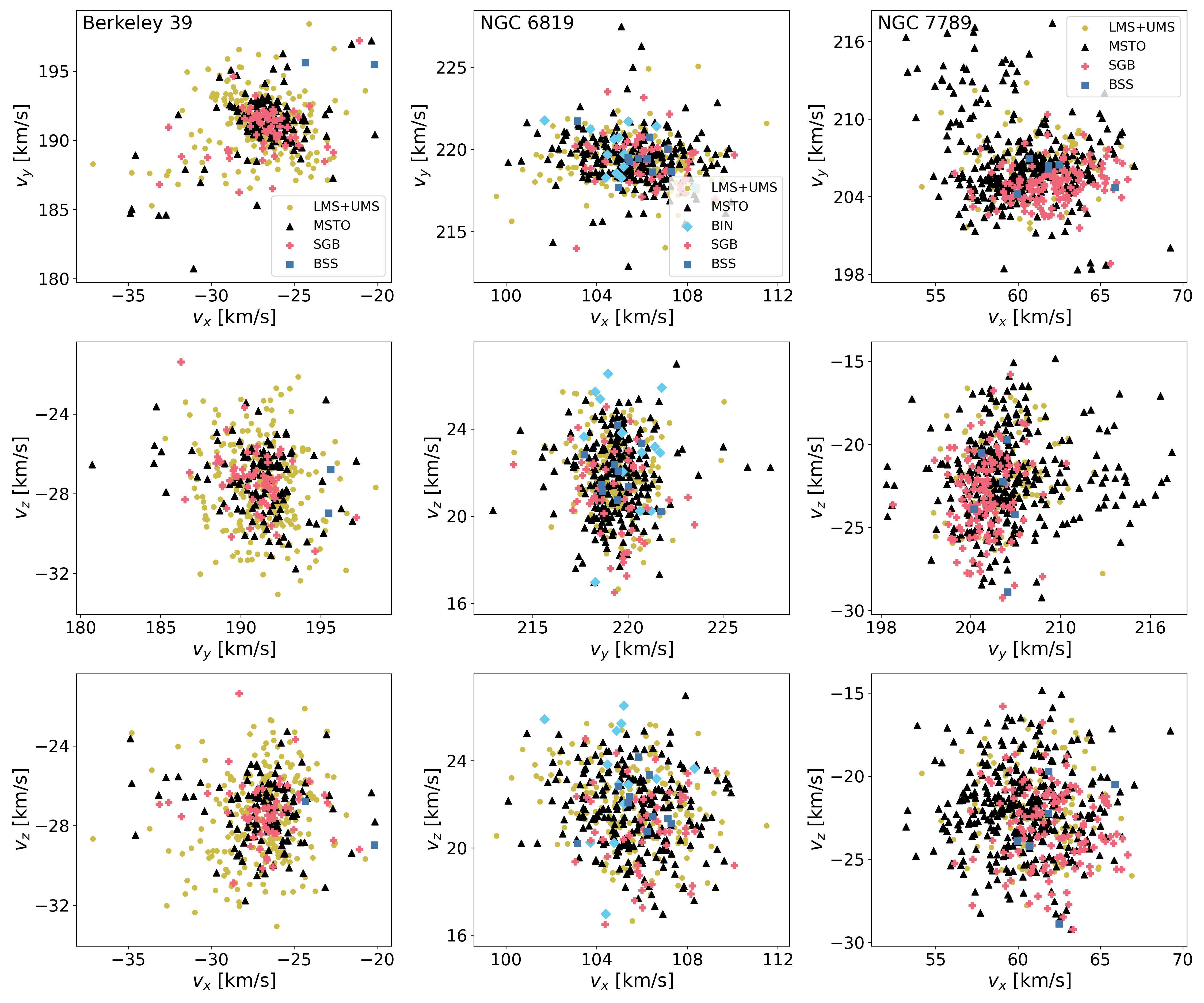}
    \caption{Velocities of the sources in Berkeley 39, NGC 6819, NGC 7789 in galactocentric coordinates, \textbf{arranged column-wise}. Different populations are represented by colored dots as per the inset legend.}
    \label{fig:galactocentric-velocities}
\end{figure*}

To further analyze the cluster dynamics, we plot the positions and velocities of different cluster populations of the four OCs in galactocentric coordinates. We seek to understand the dynamics of the heavier populations in the cluster by analyzing their positions and velocities in a 3D space. For this analysis, we combine the UMS and LMS populations into a single MS population. Since the four OCs are fairly far away, the distances from \cite{BailerJones2021} still suffer from significant line-of-sight elongation. This elongation persists in all three projections of galactocentric coordinates. Therefore, we approximate every source in the cluster to be equidistant from us and take the distance of each source to be the distance to the cluster (from Table \ref{tab:fundamental_params}). Using these distances and the RA and DEC of each source, we calculate the positions in the galactocentric coordinates. As a result, we can recover features along two projections while much of the elongation was restricted to the $(x, y)$ plane. Figure \ref{fig:galactocentric-positions} shows the resulting scatter plots in galactocentric coordinates $(x, y)$, $(y, z)$, $(x, z)$ for the four OCs. We also integrate the orbit for each cluster 20 Myr forward in time following the method from \cite{Bhattacharya2021galpy, Bhattacharya2022}. For this, we use 6D information -- positions (RA, DEC), distances, proper motions in RA and DEC, and radial velocities -- of the cross-matched sources from Table \ref{tab:galactocentric_coords} and MWPotential2014 model from the Python package galpy \citep{galpy}. The grey solid lines and the brown dashed lines in each panel of Figure \ref{fig:galactocentric-positions} represent the integrated cluster orbits and line-of-sight of each cluster, respectively. From Table \ref{tab:galactocentric_coords}, we can see that Collinder 261 does not have enough number of member with radial velocity data from \textit{Gaia} DR3. Therefore, we exclude this OC from the galactocentric velocity-space analysis and only carry out this analysis for Berkeley 39, NGC 6819, and NGC 7789. We determine the galactocentric velocities of our cross-matched members using the 6D information -- positions (RA, DEC), distances, proper motions in RA and DEC, and radial velocities. Figure \ref{fig:galactocentric-velocities} shows the scatter plots of galactocentric velocities -- $(v_x, v_y)$, $(v_y, v_z)$, $(v_x, v_z)$ -- of the different populations of Berkeley 39, NGC 6819, and NGC 7789.

\begin{table}
\centering
\caption{The fraction and number (parentheses) of cross-matched sources for each poulation with radial velocity catalogs (Gaia-ESO Public Spectroscopy Survey data (GES) for Berkeley 39 and WIYN Open Cluster Survey data (WOCS) for NGC 6819 and NGC 7789).}
\label{tab:galactocentric_coords}

\begin{tabular}{cccccc}
\hline 
\hline
Population &  & Berkeley 39          & Collinder 261        & NGC 6819             & NGC 7789    \vrule depth 2ex height 4ex width 0pt\\
\hline
BSS        &  & {10.526 (2)}  & {5.555 (2)}   & {76.923 (10)} & {63.636 (7)}\\
RGB        &  & {72.727 (64)} & {37.354 (96)} & {88.135 (52)} & {85.333 (128)}\\
MST        &  & {84.563 (126)}& {1.142 (4)}   & {89.513 (239)}& {81.481 (352)}\\
MS         &  & {47.563 (283)}& {0 (0)}       & {22.209 (199)}& {6.098 (126)}\\
BIN        &  & {N/A}         & {N/A}         & {11.764 (14)} & {0 (0)}\\[4pt]
\hline
\end{tabular}
\end{table}

\section{Discussion}
\label{sec:4discussion}

We carried out a comprehensive study to explore the dynamical stages of four OCs -- Berkeley 39, Collinder 261, NGC 6819, and NGC 7789 --by utilizing different populations, such as BSS, SGBs and RGBs, MSTOs, UMS, LMS, and binaries, These clusters are all older than 1 Gyr, therefore we anticipate that two-body relaxation to be in effect within these OCs. We provide a detailed discussion on each cluster using the analysis conducted in \S \ref{sec:3analysis}.

\subsection{Berkeley 39}

From the CRDs (Figure \ref{fig:crds}) and the AD results (Table \ref{tab:ad_test}),  we can see that the BSS appear to show no discernibly different segregation to the RGBs or the MSTOs. The MSTOs appear to be more segregated than UMS, and they both appear to be more segregated than the LMS. From the proper motion scatter diagram guided by density contours as shown in Figure \ref{fig:pm_scatter}, we observe that RGBs are the most tightly bunched in the center compared to BSS and MSTOs. However, BSS and MSTOs show comparable spread in the proper motion space. We observe a similar trend for different populations of Berkeley 39 in the galactocentric position and velocity space shown in Figure \ref{fig:galactocentric-positions} and Figure \ref{fig:galactocentric-velocities}, respectively. Since there are very few BSS compared to other cluster populations, their trend is not visible in these plots. RGBs appear to be close to the central values with a lower spread than the other two prominent populations -- MSTOs and MS stars. The bulk of the MSTOs appear to have less spread than MS stars. We also observe a weak trend of large negative $v_x$ for some sources. \cite{Vaidya2020} classified Berkeley 39 as a Family I OC because of its flat BSS radial distribution normalized with RGBs of the same magnitude range. However, they speculated that it could be a Family III cluster based on its $N_{\text{relax}}$ value. \cite{EssamBe392015} also suggests that the cluster is dynamically evolved by determining that its $N_{\text{relax}} = 78.4$. \cite{Rao2023} claimed that this cluster is of intermediate dynamical age using its $A^+_{\text{rh}}$ (0.038) and $N_{\text{relax}}$ (52), where they used average massive populations of the cluster as reference populations. We expect Berkeley 39 to show mass segregation due to energy equipartition since it is among the oldest OCs of 6 Gyr age. From this analysis, it is evident that BSS and RGBs, being the most massive and evolved populations, respectively, demonstrate a higher level of segregation. Conversely, the LMS, the lightest population, shows the least segregation. Hence, we infer that Berkeley 39 exhibits signs of mass segregation in line with what is expected for its age.

\subsection{Collinder 261}

Collinder 261 is the cluster with the largest BSS population among the four OCs studied in this work. From the CRDs (Figure \ref{fig:crds}) of different cluster populations and the AD test (Table \ref{tab:ad_test}), it is evident that BSS are more segregated than any the other populations of Collinder 261. Interestingly, the BSS population shows a strange feature, i.e., the central BSS up to $0.3r_{\text{c}}$ appear to show the same segregation as other populations, while almost 80\% of the BSS population that lies outside this range is highly segregated. The RGBs appear to be more segregated in the outer regions of the cluster than the lighter populations. However, none of the other populations show any clear sign of segregation, as is also observed in the AD test. In contrast to the results shown in CRDs (Figure \ref{fig:crds}) where a bulk of the BSS population is centrally concentrated, a large spread is observed in its proper motion distribution (Figure \ref{fig:pm_scatter}). Both the RGBs and the MSTOs show a comparable spread to the BSS.

\cite{Rain2020} found that Collinder 261 has a flat BSS radial distribution. They claim that the $N_{\text{relax}}$ value is inconclusive in determining the status of mass segregation. \cite{Rao2023} classified the cluster as having intermediate dynamical age based on its relation of $A^+_{\text{rh}}$ (0.171), with $N_{\text{relax}}$ (46). The findings from the current analysis show that all the populations show the same level of segregation except for the BSS. We speculate that the strange behaviour of the BSS is more likely related to their formation mechanisms. Despite having a significantly large number of members, its compactness parameter (log($r_{\text{t}}$/$r_{\text{c}}$) = 1.08$\pm$0.05) indicates that it has a loose core, implying that the binaries were segregated in the cluster center before the BSS formation. This behaviour is similar to an unusual case observed in the OC Melotte 66, where the BSS are highly segregated in the cluster center, resembling the most evolved OCs and GCs \citep{Rao2023}, even though the cluster's dynamical and physical parameters are similar to Collinder 261. Therefore, one needs to investigate BSS formation channels for Collinder 26 to investigate the reason behind such a large segregation of BSS. Our findings indicate that Collinder 261 is still in the early phase of mass segregation.

\subsection{NGC 6819}

NGC 6819 is one of the two clusters in this study with a clearly visible binary track. According to Figure \ref{fig:crds}, we can see that each cluster population is segregated based on its mass, except for BSS. The AD test results in Table \ref{tab:ad_test} are inconclusive in determining segregation among BSS, RGBs, and MSTOs. However, the AD test does indicate that these three populations are segregated compared to UMS and LMS, and UMS is segregated compared to LMS. Moreover, according to the AD test, the binary population exhibits different segregation levels compared to BSS, RGBs, and LMS, but the comparison to MSTOs and UMS is inconclusive.

The proper motion distribution of NGC 6819 paints a different picture. In Figure \ref{fig:pm_scatter}, we see that the BSS show the least spread among not just other populations in the cluster but other clusters in the study. The presence of outliers in the distribution of the MSTOs indicates it has a larger spread than the RGBs. In the galactocentric positions plot (Figure \ref{fig:galactocentric-positions}), we observe that most of the members belonging to the heavier BSS, RGB or MSTO populations appear to be more centrally located than the comparatively lighter binary or MS population, which are scattered across the cluster. However, since we were able to match only a small fraction of the MS or binaries, these results are partially conclusive. Almost all cross-matched BSS appear to be bunched in the center of the galactocentric velocities scatter (Fig \ref{fig:galactocentric-velocities}). However, the RGB population does not appear to be as tightly bunched, showing a comparable spread to the MSTO population. The MS and binaries also show a similar spread as RGBs and MSTOs in the galactocentric velocity distribution. This could be due to the small fraction of MS and binary population having radial velocities (Table \ref{tab:galactocentric_coords}).

\cite{Vaidya2020} observe a flat radial distribution in NGC 6819, much like Berkeley 39 and classify it as a Family I or a Family II cluster. With $A^+_{\text{rh}} = 0.248$ and $N_{\text{relax}} = 49$ from \cite{Rao2023}, NGC 6819 is classified as a intermediate dynamical age cluster. \cite{Karatas2023} observe a negative slope in the mass function and obtain a value of $N_{\text{relax}} = 108$ and claim that the cluster shows mild signs of mass segregation. \cite{Zwicker2023} compared the mass distribution of the binaries in the cluster with single stars and claimed that the cluster is mass segregated. From the current analysis, we report that NGC 6819 shows moderate signs of mass segregation.

\subsection{NGC 7789}

NGC 7789 is the youngest cluster in this study and also has a clearly visible binary track. The AD test results (Table \ref{tab:ad_test}) and CRDs indicate that LMS is the least segregated compared to BSS, RGBs, MSTOs, and UMS. However, when compared to binaries, the results are inconclusive. Furthermore, binaries show the least segregation when compared to SGBs and MSTOs. The AD test results for other comparisons remain inconclusive.

In the proper motion scatter (Figure \ref{fig:pm_scatter}), we observe that the BSS show the largest spread in the study, far more than both the RGBs and MSTOs. The RGBs and MSTOs show a comparable spread with respect to each other; however, the MSTOs have more outliers than the RGBs. In Figure \ref{fig:galactocentric-positions}, we observe that sources from all populations are scattered across the cluster, showing no signs of segregation. In Figure \ref{fig:galactocentric-velocities}, we find that some of the MSTOs have a far higher value of $v_y$ along the cluster's orbit. The RGBs seem to show an opposite trend, with most of them lying in regions with lower $v_y$ and higher $v_x$. The cross-matched BSS population or the small fraction of cross-matched MS sources do not show any such bias. To investigate the reason behind this strange behaviour of MSTOs, we plotted radial velocity distribution of these stars and found that these stars have particularly higher radial velocities. When checked their errors, we did not find any strange high errors in the radial velocities. Interestingly, we find that of the 35 MSTOs which are outliers in the $v_x$ and $v_y$ plot, 30 are very rapid rotators\footnote{According to the the WIYN open cluster survey, \citet{Nine2020}, has classified stars which have projected rotational velocity greater than 120 km/s in the very rapid rotators category.}, 2 are binary members, and 3 are binary unknowns. As we can see from its CMD (Figure \ref{fig:cmds}) this cluster falls among the cluster with extended main-sequence turnoff. It is typical for NGC 7789 to contain a large number of rapid rotators among MSTOs due to its age (1.6 Gyr). As these stars are more massive than 1.3 solar masses, they possess radiative envelopes, hence allowing them to maintain the rapid rate of rotation since their formation.

\citet{WuNGC77892007} estimated the concentration parameter and obtained a relaxation time of $516.8$ Myr for this cluster. \cite{Rao2023} obtained $N_{\text{relax}} = 4.23$ and $A^+_{\text{rh}} = 0.106$. Both claim that the cluster is of intermediate dynamical age. Based on our analyses, we find that NGC 7789 shows almost no signs of mass segregation and classify it as a dynamically young cluster.
\section{Summary and Conclusions}
\label{sec:5conclusions}

In this study, we explored the degree of mass segregation of the four OCs -- Berkeley 39, Collinder 261, NGC 6819, and NGC 7789 -- by examining the kinematics and spatial properties of their members.  We used the membership determination algorithm, ML-MOC, on \textit{Gaia} DR3 data to identify members of the four OCs. To conduct this study, we also used radial velocity data from large public spectroscopic surveys, such as GES \cite{Bragaglia2022} and WOCS \cite{WOCS2000}, in addition to \textit{Gaia} positions and proper motions.  We classified the members into different populations based on their photometric properties, including BSS, RGBs, MSTOs, UMS, and LMS, and included binaries as an additional population for NGC 6819 and NGC 7789.

Our investigations revealed that NGC 6819 and Berkeley 39 show moderate signs of mass segregation. On the other hand, both Collinder 261 and NGC 7789 exhibit fewer signs of mass segregation. Nonetheless, both clusters demonstrate peculiar characteristics. The BSS population of Collinder 261 is highly segregated in the cluster center compared to any other population. In OCs, BSS usually forms through binary system evolution, especially mass transfer. Therefore, we speculate that this could be due to the loose core of the cluster, where binaries are segregated in the cluster center prior to the BSS formation. However, further exploration of the BSS population would be needed to confirm their formation and unusually higher segregation in the cluster center. Additionally, NGC 7789 has MSTOs with unusually high galactocentric velocities, particularly in v$_y$. We found that these stars are situated at the higher end of the radial velocity distribution, and 30 out of 35 are rapid rotators ($v$sin$i > 120$ km/s). We speculate that the large radial velocities of these stars could be associated with their rapid rotation. Despite the varying degrees of mass segregation observed in these OCs, none of them are fully evolved. As OCs reside in the galactic plane, it is expected that they eventually disintegrate before reaching a fully mass segregated state. While $A^+_{\text{rh}}$ and $N_{\text{relax}}$ serve as good indicators for dynamical evolution when comparing among a large sample of OCs, individual systems may show deviations from the derived cluster dynamical properties based on $A^+_{\text{rh}}$ and $N_{\text{relax}}$. This is likely due to higher galactic interactions experienced by OCs as they are generally in the disc. Therefore, a more holistic analysis as carried out here is required for better understanding the dynamical state of OCs. By analyzing the cluster populations separately, we have attempted to understand how they represent the overall mass segregation of these four OCs and thus their dynamical status.

\section*{Acknowledgements}

We thank the referee for their comments that helped improve the manuscript. KKR acknowledges the financial support from National Science and Technology Council, Taiwan, R.O.C. (NSTC 113-2811-M-008-034). SB is funded by the INSPIRE Faculty Award (DST/INSPIRE/04/2020/002224), Department of Science and Technology, Government of India. This work has made use of the early third data release from the European Space Agency mission Gaia (\url{https://www.cosmos.esa.int/gaia}), Gaia DR3 \citep{Gaia2023}, processed by the Gaia Data Processing and Analysis Consortium (DPAC; \url{https://www.cosmos.esa.int/web/gaia/dpac/consortium}). Funding for the DPAC has been provided by national institutions, in particular, the institutions participating in the Gaia Multilateral Agreement. This research has made use of the Vizier catalogue access tool, CDS, Strasbourg, France. This research made use of ASTROPY, a PYTHON package for astronomy \citep{astropy}, NUMPY \citep{NumPy}, MATPLOTLIB \citep{Matplotlib}, and SCIPY \citep{SciPy}. This research also made use of the Astrophysics Data System governed by NASA (\url{https://ui.adsabs.harvard.edu}).

\bibliography{references}{}

\begin{thebibliography}{}
\expandafter\ifx\csname natexlab\endcsname\relax\def\natexlab#1{#1}\fi
\providecommand{\url}[1]{\href{#1}{#1}}
\providecommand{\dodoi}[1]{doi:~\href{http://doi.org/#1}{\nolinkurl{#1}}}
\providecommand{\doeprint}[1]{\href{http://ascl.net/#1}{\nolinkurl{http://ascl.net/#1}}}
\providecommand{\doarXiv}[1]{\href{https://arxiv.org/abs/#1}{\nolinkurl{https://arxiv.org/abs/#1}}}

\bibitem[{{Agarwal} {et~al.}(2021){Agarwal}, {Rao}, {Vaidya}, \& {Bhattacharya}}]{Agarwal2021}
{Agarwal}, M., {Rao}, K.~K., {Vaidya}, K., \& {Bhattacharya}, S. 2021, \mnras, 502, 2582, \dodoi{10.1093/mnras/stab118}

\bibitem[{{Alessandrini} {et~al.}(2016){Alessandrini}, {Lanzoni}, {Ferraro}, {Miocchi}, \& {Vesperini}}]{Alessandrini2016}
{Alessandrini}, E., {Lanzoni}, B., {Ferraro}, F.~R., {Miocchi}, P., \& {Vesperini}, E. 2016, \apj, 833, 252, \dodoi{10.3847/1538-4357/833/2/252}

\bibitem[{{Astropy Collaboration} {et~al.}(2022){Astropy Collaboration}, {Price-Whelan}, {Lim}, {Earl}, {Starkman}, {Bradley}, {Shupe}, {Patil}, {Corrales}, {Brasseur}, {N{"o}the}, {Donath}, {Tollerud}, {Morris}, {Ginsburg}, {Vaher}, {Weaver}, {Tocknell}, {Jamieson}, {van Kerkwijk}, {Robitaille}, {Merry}, {Bachetti}, {G{"u}nther}, {Aldcroft}, {Alvarado-Montes}, {Archibald}, {B{'o}di}, {Bapat}, {Barentsen}, {Baz{'a}n}, {Biswas}, {Boquien}, {Burke}, {Cara}, {Cara}, {Conroy}, {Conseil}, {Craig}, {Cross}, {Cruz}, {D'Eugenio}, {Dencheva}, {Devillepoix}, {Dietrich}, {Eigenbrot}, {Erben}, {Ferreira}, {Foreman-Mackey}, {Fox}, {Freij}, {Garg}, {Geda}, {Glattly}, {Gondhalekar}, {Gordon}, {Grant}, {Greenfield}, {Groener}, {Guest}, {Gurovich}, {Handberg}, {Hart}, {Hatfield-Dodds}, {Homeier}, {Hosseinzadeh}, {Jenness}, {Jones}, {Joseph}, {Kalmbach}, {Karamehmetoglu}, {Ka{l}uszy{'n}ski}, {Kelley}, {Kern}, {Kerzendorf}, {Koch}, {Kulumani}, {Lee}, {Ly}, {Ma}, {MacBride}, {Maljaars}, {Muna}, {Murphy}, {Norman}, {O'Steen},
  {Oman}, {Pacifici}, {Pascual}, {Pascual-Granado}, {Patil}, {Perren}, {Pickering}, {Rastogi}, {Roulston}, {Ryan}, {Rykoff}, {Sabater}, {Sakurikar}, {Salgado}, {Sanghi}, {Saunders}, {Savchenko}, {Schwardt}, {Seifert-Eckert}, {Shih}, {Jain}, {Shukla}, {Sick}, {Simpson}, {Singanamalla}, {Singer}, {Singhal}, {Sinha}, {Sip{H{o}}cz}, {Spitler}, {Stansby}, {Streicher}, {{{S}}umak}, {Swinbank}, {Taranu}, {Tewary}, {Tremblay}, {Val-Borro}, {Van Kooten}, {Vasovi{'c}}, {Verma}, {de Miranda Cardoso}, {Williams}, {Wilson}, {Winkel}, {Wood-Vasey}, {Xue}, {Yoachim}, {Zhang}, {Zonca}, \& {Astropy Project Contributors}}]{astropy}
{Astropy Collaboration}, {Price-Whelan}, A.~M., {Lim}, P.~L., {et~al.} 2022, apj, 935, 167, \dodoi{10.3847/1538-4357/ac7c74}

\bibitem[{{Bailer-Jones} {et~al.}(2021){Bailer-Jones}, {Rybizki}, {Fouesneau}, {Demleitner}, \& {Andrae}}]{BailerJones2021}
{Bailer-Jones}, C.~A.~L., {Rybizki}, J., {Fouesneau}, M., {Demleitner}, M., \& {Andrae}, R. 2021, VizieR Online Data Catalog, I/352

\bibitem[{{Bailyn}(1995)}]{Bailyn1995}
{Bailyn}, C.~D. 1995, \araa, 33, 133, \dodoi{10.1146/annurev.aa.33.090195.001025}

\bibitem[{{Baumgardt} \& {Makino}(2003)}]{Baumgardt2003}
{Baumgardt}, H., \& {Makino}, J. 2003, \mnras, 340, 227, \dodoi{10.1046/j.1365-8711.2003.06286.x}

\bibitem[{{Beccari} {et~al.}(2023){Beccari}, {Cadelano}, \& {Dalessandro}}]{Beccari2023}
{Beccari}, G., {Cadelano}, M., \& {Dalessandro}, E. 2023, \aap, 670, A11, \dodoi{10.1051/0004-6361/202244288}

\bibitem[{{Beccari} {et~al.}(2013){Beccari}, {Dalessandro}, {Lanzoni}, {Ferraro}, {Sollima}, {Bellazzini}, \& {Miocchi}}]{Beccari2013}
{Beccari}, G., {Dalessandro}, E., {Lanzoni}, B., {et~al.} 2013, \apj, 776, 60, \dodoi{10.1088/0004-637X/776/1/60}

\bibitem[{{Bhattacharya} {et~al.}(2021{\natexlab{a}}){Bhattacharya}, {Agarwal}, {Rao}, \& {Vaidya}}]{Bhattacharya2021galpy}
{Bhattacharya}, S., {Agarwal}, M., {Rao}, K.~K., \& {Vaidya}, K. 2021{\natexlab{a}}, \mnras, 505, 1607, \dodoi{10.1093/mnras/stab1404}

\bibitem[{{Bhattacharya} {et~al.}(2021{\natexlab{b}}){Bhattacharya}, {Arnaboldi}, {Gerhard}, {McConnachie}, {Caldwell}, {Hartke}, \& {Freeman}}]{Bhattacharya2021ADTest}
{Bhattacharya}, S., {Arnaboldi}, M., {Gerhard}, O., {et~al.} 2021{\natexlab{b}}, \aap, 647, A130, \dodoi{10.1051/0004-6361/202038366}

\bibitem[{{Bhattacharya} {et~al.}(2022){Bhattacharya}, {Rao}, {Agarwal}, {Balan}, \& {Vaidya}}]{Bhattacharya2022}
{Bhattacharya}, S., {Rao}, K.~K., {Agarwal}, M., {Balan}, S., \& {Vaidya}, K. 2022, \mnras, 517, 3525, \dodoi{10.1093/mnras/stac2906}

\bibitem[{{Bhattacharya} {et~al.}(2019){Bhattacharya}, {Vaidya, Kaushar}, {Chen, W. P.}, \& {Beccari, Giacomo}}]{Bhattacharya2019}
{Bhattacharya}, S., {Vaidya, Kaushar}, {Chen, W. P.}, \& {Beccari, Giacomo}. 2019, A\&A, 624, A26, \dodoi{10.1051/0004-6361/201834449}

\bibitem[{Bianchini {et~al.}(2016)Bianchini, Bianchini, van~de Ven, van~de Ven, Norris, Schinnerer, \& Varri}]{Bianchini2016}
Bianchini, P., Bianchini, P., van~de Ven, G., {et~al.} 2016, Monthly Notices of the Royal Astronomical Society, \dodoi{10.1093/mnras/stw552}

\bibitem[{Bovy(2015)}]{galpy}
Bovy, J. 2015, Astrophys. J. Supp., 29, 216

\bibitem[{{Bragaglia} {et~al.}(2022){Bragaglia}, {Alfaro}, {Flaccomio}, {Blomme}, {Donati}, {Costado}, {Damiani}, {Franciosini}, {Prisinzano}, {Randich}, {Friel}, {Hatztidimitriou}, {Vallenari}, {Spagna}, {Balaguer-Nunez}, {Bonito}, {Cantat Gaudin}, {Casamiquela}, {Jeffries}, {Jordi}, {Magrini}, {Drew}, {Jackson}, {Abbas}, {Caramazza}, {Hayes}, {Jim{\'e}nez-Esteban}, {Re Fiorentin}, {Wright}, {Bayo}, {Bensby}, {Bergemann}, {Gilmore}, {Gonneau}, {Heiter}, {Hourihane}, {Pancino}, {Sacco}, {Smiljanic}, {Zaggia}, \& {Vink}}]{Bragaglia2022}
{Bragaglia}, A., {Alfaro}, E.~J., {Flaccomio}, E., {et~al.} 2022, \aap, 659, A200, \dodoi{10.1051/0004-6361/202142674}

\bibitem[{Bressan {et~al.}(2012)Bressan, Marigo, Girardi, Salasnich, Dal~Cero, Rubele, \& Nanni}]{PARSEC_Isochrones}
Bressan, A., Marigo, P., Girardi, L., {et~al.} 2012, Monthly Notices of the Royal Astronomical Society, 427, 127, \dodoi{10.1111/j.1365-2966.2012.21948.x}

\bibitem[{{Cadelano} {et~al.}(2022){Cadelano}, {Ferraro}, {Dalessandro}, {Lanzoni}, {Pallanca}, \& {Saracino}}]{Cadelano2022}
{Cadelano}, M., {Ferraro}, F.~R., {Dalessandro}, E., {et~al.} 2022, \apj, 941, 69, \dodoi{10.3847/1538-4357/aca016}

\bibitem[{{Cantat-Gaudin}(2022)}]{Cantat-Gaudin2022}
{Cantat-Gaudin}, T. 2022, Universe, 8, 111, \dodoi{10.3390/universe8020111}

\bibitem[{{Chandrasekhar}(1943)}]{Chandrasekhar1943}
{Chandrasekhar}, S. 1943, \apj, 97, 263, \dodoi{10.1086/144518}

\bibitem[{Comaniciu \& Meer(2002)}]{ComaniciuMeer2002}
Comaniciu, D., \& Meer, P. 2002, IEEE Transactions on Pattern Analysis and Machine Intelligence, 24, 603, \dodoi{10.1109/34.1000236}

\bibitem[{Cover \& Hart(1967)}]{CoverHart1967}
Cover, T., \& Hart, P. 1967, IEEE Transactions on Information Theory, 13, 21, \dodoi{10.1109/TIT.1967.1053964}

\bibitem[{{Dalessandro} {et~al.}(2015){Dalessandro}, {Ferraro}, {Massari}, {Lanzoni}, {Miocchi}, \& {Beccari}}]{Dalessandro2015}
{Dalessandro}, E., {Ferraro}, F.~R., {Massari}, D., {et~al.} 2015, \apj, 810, 40, \dodoi{10.1088/0004-637X/810/1/40}

\bibitem[{{Essam} \& {Selim}(2015)}]{EssamBe392015}
{Essam}, A., \& {Selim}, I.~M. 2015, International Journal of Astronomy and Astrophysics, 5, 173, \dodoi{10.4236/ijaa.2015.53022}

\bibitem[{{Ferraro} {et~al.}(2020){Ferraro}, {Lanzoni}, \& {Dalessandro}}]{Ferraro2020}
{Ferraro}, F.~R., {Lanzoni}, B., \& {Dalessandro}, E. 2020, Rendiconti Lincei. Scienze Fisiche e Naturali, 31, 19, \dodoi{10.1007/s12210-020-00873-2}

\bibitem[{{Ferraro} {et~al.}(2012){Ferraro}, {Lanzoni}, {Dalessandro}, {Beccari}, {Pasquato}, {Miocchi}, {Rood}, {Sigurdsson}, {Sills}, {Vesperini}, {Mapelli}, {Contreras}, {Sanna}, \& {Mucciarelli}}]{Ferraro2012}
{Ferraro}, F.~R., {Lanzoni}, B., {Dalessandro}, E., {et~al.} 2012, \nat, 492, 393, \dodoi{10.1038/nature11686}

\bibitem[{{Gaia Collaboration} {et~al.}(2016){Gaia Collaboration}, {Prusti}, {de Bruijne}, {Brown}, {Vallenari}, {Babusiaux}, {Bailer-Jones}, {Bastian}, {Biermann}, {Evans}, {Eyer}, {Jansen}, {Jordi}, {Klioner}, {Lammers}, {Lindegren}, {Luri}, {Mignard}, {Milligan}, {Panem}, {Poinsignon}, {Pourbaix}, {Randich}, {Sarri}, {Sartoretti}, {Siddiqui}, {Soubiran}, {Valette}, {van Leeuwen}, {Walton}, {Aerts}, {Arenou}, {Cropper}, {Drimmel}, {H{\o}g}, {Katz}, {Lattanzi}, {O'Mullane}, {Grebel}, {Holland}, {Huc}, {Passot}, {Bramante}, {Cacciari}, {Casta{\~n}eda}, {Chaoul}, {Cheek}, {De Angeli}, {Fabricius}, {Guerra}, {Hern{\'a}ndez}, {Jean-Antoine-Piccolo}, {Masana}, {Messineo}, {Mowlavi}, {Nienartowicz}, {Ord{\'o}{\~n}ez-Blanco}, {Panuzzo}, {Portell}, {Richards}, {Riello}, {Seabroke}, {Tanga}, {Th{\'e}venin}, {Torra}, {Els}, {Gracia-Abril}, {Comoretto}, {Garcia-Reinaldos}, {Lock}, {Mercier}, {Altmann}, {Andrae}, {Astraatmadja}, {Bellas-Velidis}, {Benson}, {Berthier}, {Blomme}, {Busso}, {Carry}, {Cellino}, {Clementini},
  {Cowell}, {Creevey}, {Cuypers}, {Davidson}, {De Ridder}, {de Torres}, {Delchambre}, {Dell'Oro}, {Ducourant}, {Fr{\'e}mat}, {Garc{\'\i}a-Torres}, {Gosset}, {Halbwachs}, {Hambly}, {Harrison}, {Hauser}, {Hestroffer}, {Hodgkin}, {Huckle}, {Hutton}, {Jasniewicz}, {Jordan}, {Kontizas}, {Korn}, {Lanzafame}, {Manteiga}, {Moitinho}, {Muinonen}, {Osinde}, {Pancino}, {Pauwels}, {Petit}, {Recio-Blanco}, {Robin}, {Sarro}, {Siopis}, {Smith}, {Smith}, {Sozzetti}, {Thuillot}, {van Reeven}, {Viala}, {Abbas}, {Abreu Aramburu}, {Accart}, {Aguado}, {Allan}, {Allasia}, {Altavilla}, {{\'A}lvarez}, {Alves}, {Anderson}, {Andrei}, {Anglada Varela}, {Antiche}, {Antoja}, {Ant{\'o}n}, {Arcay}, {Atzei}, {Ayache}, {Bach}, {Baker}, {Balaguer-N{\'u}{\~n}ez}, {Barache}, {Barata}, {Barbier}, {Barblan}, {Baroni}, {Barrado y Navascu{\'e}s}, {Barros}, {Barstow}, {Becciani}, {Bellazzini}, {Bellei}, {Bello Garc{\'\i}a}, {Belokurov}, {Bendjoya}, {Berihuete}, {Bianchi}, {Bienaym{\'e}}, {Billebaud}, {Blagorodnova}, {Blanco-Cuaresma}, {Boch},
  {Bombrun}, {Borrachero}, {Bouquillon}, {Bourda}, {Bouy}, {Bragaglia}, {Breddels}, {Brouillet}, {Br{\"u}semeister}, {Bucciarelli}, {Budnik}, {Burgess}, {Burgon}, {Burlacu}, {Busonero}, {Buzzi}, {Caffau}, {Cambras}, {Campbell}, {Cancelliere}, {Cantat-Gaudin}, {Carlucci}, {Carrasco}, {Castellani}, {Charlot}, {Charnas}, {Charvet}, {Chassat}, {Chiavassa}, {Clotet}, {Cocozza}, {Collins}, {Collins}, {Costigan}, {Crifo}, {Cross}, {Crosta}, {Crowley}, {Dafonte}, {Damerdji}, {Dapergolas}, {David}, {David}, {De Cat}, {de Felice}, {de Laverny}, {De Luise}, {De March}, {de Martino}, {de Souza}, {Debosscher}, {del Pozo}, {Delbo}, {Delgado}, {Delgado}, {di Marco}, {Di Matteo}, {Diakite}, {Distefano}, {Dolding}, {Dos Anjos}, {Drazinos}, {Dur{\'a}n}, {Dzigan}, {Ecale}, {Edvardsson}, {Enke}, {Erdmann}, {Escolar}, {Espina}, {Evans}, {Eynard Bontemps}, {Fabre}, {Fabrizio}, {Faigler}, {Falc{\~a}o}, {Farr{\`a}s Casas}, {Faye}, {Federici}, {Fedorets}, {Fern{\'a}ndez-Hern{\'a}ndez}, {Fernique}, {Fienga}, {Figueras}, {Filippi},
  {Findeisen}, {Fonti}, {Fouesneau}, {Fraile}, {Fraser}, {Fuchs}, {Furnell}, {Gai}, {Galleti}, {Galluccio}, {Garabato}, {Garc{\'\i}a-Sedano}, {Gar{\'e}}, {Garofalo}, {Garralda}, {Gavras}, {Gerssen}, {Geyer}, {Gilmore}, {Girona}, {Giuffrida}, {Gomes}, {Gonz{\'a}lez-Marcos}, {Gonz{\'a}lez-N{\'u}{\~n}ez}, {Gonz{\'a}lez-Vidal}, {Granvik}, {Guerrier}, {Guillout}, {Guiraud}, {G{\'u}rpide}, {Guti{\'e}rrez-S{\'a}nchez}, {Guy}, {Haigron}, {Hatzidimitriou}, {Haywood}, {Heiter}, {Helmi}, {Hobbs}, {Hofmann}, {Holl}, {Holland}, {Hunt}, {Hypki}, {Icardi}, {Irwin}, {Jevardat de Fombelle}, {Jofr{\'e}}, {Jonker}, {Jorissen}, {Julbe}, {Karampelas}, {Kochoska}, {Kohley}, {Kolenberg}, {Kontizas}, {Koposov}, {Kordopatis}, {Koubsky}, {Kowalczyk}, {Krone-Martins}, {Kudryashova}, {Kull}, {Bachchan}, {Lacoste-Seris}, {Lanza}, {Lavigne}, {Le Poncin-Lafitte}, {Lebreton}, {Lebzelter}, {Leccia}, {Leclerc}, {Lecoeur-Taibi}, {Lemaitre}, {Lenhardt}, {Leroux}, {Liao}, {Licata}, {Lindstr{\o}m}, {Lister}, {Livanou}, {Lobel}, {L{\"o}ffler},
  {L{\'o}pez}, {Lopez-Lozano}, {Lorenz}, {Loureiro}, {MacDonald}, {Magalh{\~a}es Fernandes}, {Managau}, {Mann}, {Mantelet}, {Marchal}, {Marchant}, {Marconi}, {Marie}, {Marinoni}, {Marrese}, {Marschalk{\'o}}, {Marshall}, {Mart{\'\i}n-Fleitas}, {Martino}, {Mary}, {Matijevi{\v{c}}}, {Mazeh}, {McMillan}, {Messina}, {Mestre}, {Michalik}, {Millar}, {Miranda}, {Molina}, {Molinaro}, {Molinaro}, {Moln{\'a}r}, {Moniez}, {Montegriffo}, {Monteiro}, {Mor}, {Mora}, {Morbidelli}, {Morel}, {Morgenthaler}, {Morley}, {Morris}, {Mulone}, {Muraveva}, {Musella}, {Narbonne}, {Nelemans}, {Nicastro}, {Noval}, {Ord{\'e}novic}, {Ordieres-Mer{\'e}}, {Osborne}, {Pagani}, {Pagano}, {Pailler}, {Palacin}, {Palaversa}, {Parsons}, {Paulsen}, {Pecoraro}, {Pedrosa}, {Pentik{\"a}inen}, {Pereira}, {Pichon}, {Piersimoni}, {Pineau}, {Plachy}, {Plum}, {Poujoulet}, {Pr{\v{s}}a}, {Pulone}, {Ragaini}, {Rago}, {Rambaux}, {Ramos-Lerate}, {Ranalli}, {Rauw}, {Read}, {Regibo}, {Renk}, {Reyl{\'e}}, {Ribeiro}, {Rimoldini}, {Ripepi}, {Riva}, {Rixon},
  {Roelens}, {Romero-G{\'o}mez}, {Rowell}, {Royer}, {Rudolph}, {Ruiz-Dern}, {Sadowski}, {Sagrist{\`a} Sell{\'e}s}, {Sahlmann}, {Salgado}, {Salguero}, {Sarasso}, {Savietto}, {Schnorhk}, {Schultheis}, {Sciacca}, {Segol}, {Segovia}, {Segransan}, {Serpell}, {Shih}, {Smareglia}, {Smart}, {Smith}, {Solano}, {Solitro}, {Sordo}, {Soria Nieto}, {Souchay}, {Spagna}, {Spoto}, {Stampa}, {Steele}, {Steidelm{\"u}ller}, {Stephenson}, {Stoev}, {Suess}, {S{\"u}veges}, {Surdej}, {Szabados}, {Szegedi-Elek}, {Tapiador}, {Taris}, {Tauran}, {Taylor}, {Teixeira}, {Terrett}, {Tingley}, {Trager}, {Turon}, {Ulla}, {Utrilla}, {Valentini}, {van Elteren}, {Van Hemelryck}, {van Leeuwen}, {Varadi}, {Vecchiato}, {Veljanoski}, {Via}, {Vicente}, {Vogt}, {Voss}, {Votruba}, {Voutsinas}, {Walmsley}, {Weiler}, {Weingrill}, {Werner}, {Wevers}, {Whitehead}, {Wyrzykowski}, {Yoldas}, {{\v{Z}}erjal}, {Zucker}, {Zurbach}, {Zwitter}, {Alecu}, {Allen}, {Allende Prieto}, {Amorim}, {Anglada-Escud{\'e}}, {Arsenijevic}, {Azaz}, {Balm}, {Beck}, {Bernstein},
  {Bigot}, {Bijaoui}, {Blasco}, {Bonfigli}, {Bono}, {Boudreault}, {Bressan}, {Brown}, {Brunet}, {Bunclark}, {Buonanno}, {Butkevich}, {Carret}, {Carrion}, {Chemin}, {Ch{\'e}reau}, {Corcione}, {Darmigny}, {de Boer}, {de Teodoro}, {de Zeeuw}, {Delle Luche}, {Domingues}, {Dubath}, {Fodor}, {Fr{\'e}zouls}, {Fries}, {Fustes}, {Fyfe}, {Gallardo}, {Gallegos}, {Gardiol}, {Gebran}, {Gomboc}, {G{\'o}mez}, {Grux}, {Gueguen}, {Heyrovsky}, {Hoar}, {Iannicola}, {Isasi Parache}, {Janotto}, {Joliet}, {Jonckheere}, {Keil}, {Kim}, {Klagyivik}, {Klar}, {Knude}, {Kochukhov}, {Kolka}, {Kos}, {Kutka}, {Lainey}, {LeBouquin}, {Liu}, {Loreggia}, {Makarov}, {Marseille}, {Martayan}, {Martinez-Rubi}, {Massart}, {Meynadier}, {Mignot}, {Munari}, {Nguyen}, {Nordlander}, {Ocvirk}, {O'Flaherty}, {Olias Sanz}, {Ortiz}, {Osorio}, {Oszkiewicz}, {Ouzounis}, {Palmer}, {Park}, {Pasquato}, {Peltzer}, {Peralta}, {P{\'e}turaud}, {Pieniluoma}, {Pigozzi}, {Poels}, {Prat}, {Prod'homme}, {Raison}, {Rebordao}, {Risquez}, {Rocca-Volmerange}, {Rosen},
  {Ruiz-Fuertes}, {Russo}, {Sembay}, {Serraller Vizcaino}, {Short}, {Siebert}, {Silva}, {Sinachopoulos}, {Slezak}, {Soffel}, {Sosnowska}, {Strai{\v{z}}ys}, {ter Linden}, {Terrell}, {Theil}, {Tiede}, {Troisi}, {Tsalmantza}, {Tur}, {Vaccari}, {Vachier}, {Valles}, {Van Hamme}, {Veltz}, {Virtanen}, {Wallut}, {Wichmann}, {Wilkinson}, {Ziaeepour}, \& {Zschocke}}]{Gaia2016}
{Gaia Collaboration}, {Prusti}, T., {de Bruijne}, J.~H.~J., {et~al.} 2016, \aap, 595, A1, \dodoi{10.1051/0004-6361/201629272}

\bibitem[{{Gaia Collaboration} {et~al.}(2023){Gaia Collaboration}, {Vallenari, A.}, {Brown, A. G. A.}, {Prusti, T.}, {de Bruijne, J. H. J.}, {Arenou, F.}, {Babusiaux, C.}, {Biermann, M.}, {Creevey, O. L.}, {Ducourant, C.}, {Evans, D. W.}, {Eyer, L.}, {Guerra, R.}, {Hutton, A.}, {Jordi, C.}, {Klioner, S. A.}, {Lammers, U. L.}, {Lindegren, L.}, {Luri, X.}, {Mignard, F.}, {Panem, C.}, {Pourbaix, D.}, {Randich, S.}, {Sartoretti, P.}, {Soubiran, C.}, {Tanga, P.}, {Walton, N. A.}, {Bailer-Jones, C. A. L.}, {Bastian, U.}, {Drimmel, R.}, {Jansen, F.}, {Katz, D.}, {Lattanzi, M. G.}, {van Leeuwen, F.}, {Bakker, J.}, {Cacciari, C.}, {Casta\~neda, J.}, {De Angeli, F.}, {Fabricius, C.}, {Fouesneau, M.}, {Fr\'emat, Y.}, {Galluccio, L.}, {Guerrier, A.}, {Heiter, U.}, {Masana, E.}, {Messineo, R.}, {Mowlavi, N.}, {Nicolas, C.}, {Nienartowicz, K.}, {Pailler, F.}, {Panuzzo, P.}, {Riclet, F.}, {Roux, W.}, {Seabroke, G. M.}, {Sordo, R.}, {Th\'evenin, F.}, {Gracia-Abril, G.}, {Portell, J.}, {Teyssier, D.}, {Altmann, M.}, {Andrae,
  R.}, {Audard, M.}, {Bellas-Velidis, I.}, {Benson, K.}, {Berthier, J.}, {Blomme, R.}, {Burgess, P. W.}, {Busonero, D.}, {Busso, G.}, {C\'anovas, H.}, {Carry, B.}, {Cellino, A.}, {Cheek, N.}, {Clementini, G.}, {Damerdji, Y.}, {Davidson, M.}, {de Teodoro, P.}, {Nu\~nez Campos, M.}, {Delchambre, L.}, {Dell\'{}Oro, A.}, {Esquej, P.}, {Fern\'andez-Hern\'andez, J.}, {Fraile, E.}, {Garabato, D.}, {Garc\'{\i}a-Lario, P.}, {Gosset, E.}, {Haigron, R.}, {Halbwachs, J.-L.}, {Hambly, N. C.}, {Harrison, D. L.}, {Hern\'andez, J.}, {Hestroffer, D.}, {Hodgkin, S. T.}, {Holl, B.}, {Jan\ss{}en, K.}, {Jevardat de Fombelle, G.}, {Jordan, S.}, {Krone-Martins, A.}, {Lanzafame, A. C.}, {L\"offler, W.}, {Marchal, O.}, {Marrese, P. M.}, {Moitinho, A.}, {Muinonen, K.}, {Osborne, P.}, {Pancino, E.}, {Pauwels, T.}, {Recio-Blanco, A.}, {Reyl\'e, C.}, {Riello, M.}, {Rimoldini, L.}, {Roegiers, T.}, {Rybizki, J.}, {Sarro, L. M.}, {Siopis, C.}, {Smith, M.}, {Sozzetti, A.}, {Utrilla, E.}, {van Leeuwen, M.}, {Abbas, U.}, {\'Abrah\'am, P.},
  {Abreu Aramburu, A.}, {Aerts, C.}, {Aguado, J. J.}, {Ajaj, M.}, {Aldea-Montero, F.}, {Altavilla, G.}, {\'Alvarez, M. A.}, {Alves, J.}, {Anders, F.}, {Anderson, R. I.}, {Anglada Varela, E.}, {Antoja, T.}, {Baines, D.}, {Baker, S. G.}, {Balaguer-N\'u\~nez, L.}, {Balbinot, E.}, {Balog, Z.}, {Barache, C.}, {Barbato, D.}, {Barros, M.}, {Barstow, M. A.}, {Bartolom\'e, S.}, {Bassilana, J.-L.}, {Bauchet, N.}, {Becciani, U.}, {Bellazzini, M.}, {Berihuete, A.}, {Bernet, M.}, {Bertone, S.}, {Bianchi, L.}, {Binnenfeld, A.}, {Blanco-Cuaresma, S.}, {Blazere, A.}, {Boch, T.}, {Bombrun, A.}, {Bossini, D.}, {Bouquillon, S.}, {Bragaglia, A.}, {Bramante, L.}, {Breedt, E.}, {Bressan, A.}, {Brouillet, N.}, {Brugaletta, E.}, {Bucciarelli, B.}, {Burlacu, A.}, {Butkevich, A. G.}, {Buzzi, R.}, {Caffau, E.}, {Cancelliere, R.}, {Cantat-Gaudin, T.}, {Carballo, R.}, {Carlucci, T.}, {Carnerero, M. I.}, {Carrasco, J. M.}, {Casamiquela, L.}, {Castellani, M.}, {Castro-Ginard, A.}, {Chaoul, L.}, {Charlot, P.}, {Chemin, L.}, {Chiaramida,
  V.}, {Chiavassa, A.}, {Chornay, N.}, {Comoretto, G.}, {Contursi, G.}, {Cooper, W. J.}, {Cornez, T.}, {Cowell, S.}, {Crifo, F.}, {Cropper, M.}, {Crosta, M.}, {Crowley, C.}, {Dafonte, C.}, {Dapergolas, A.}, {David, M.}, {David, P.}, {de Laverny, P.}, {De Luise, F.}, {De March, R.}, {De Ridder, J.}, {de Souza, R.}, {de Torres, A.}, {del Peloso, E. F.}, {del Pozo, E.}, {Delbo, M.}, {Delgado, A.}, {Delisle, J.-B.}, {Demouchy, C.}, {Dharmawardena, T. E.}, {Di Matteo, P.}, {Diakite, S.}, {Diener, C.}, {Distefano, E.}, {Dolding, C.}, {Edvardsson, B.}, {Enke, H.}, {Fabre, C.}, {Fabrizio, M.}, {Faigler, S.}, {Fedorets, G.}, {Fernique, P.}, {Fienga, A.}, {Figueras, F.}, {Fournier, Y.}, {Fouron, C.}, {Fragkoudi, F.}, {Gai, M.}, {Garcia-Gutierrez, A.}, {Garcia-Reinaldos, M.}, {Garc\'{\i}a-Torres, M.}, {Garofalo, A.}, {Gavel, A.}, {Gavras, P.}, {Gerlach, E.}, {Geyer, R.}, {Giacobbe, P.}, {Gilmore, G.}, {Girona, S.}, {Giuffrida, G.}, {Gomel, R.}, {Gomez, A.}, {Gonz\'alez-N\'u\~nez, J.}, {Gonz\'alez-Santamar\'{\i}a, I.},
  {Gonz\'alez-Vidal, J. J.}, {Granvik, M.}, {Guillout, P.}, {Guiraud, J.}, {Guti\'errez-S\'anchez, R.}, {Guy, L. P.}, {Hatzidimitriou, D.}, {Hauser, M.}, {Haywood, M.}, {Helmer, A.}, {Helmi, A.}, {Sarmiento, M. H.}, {Hidalgo, S. L.}, {Hilger, T.}, {Hladczuk, N.}, {Hobbs, D.}, {Holland, G.}, {Huckle, H. E.}, {Jardine, K.}, {Jasniewicz, G.}, {Jean-Antoine Piccolo, A.}, {Jim\'enez-Arranz, \'O.}, {Jorissen, A.}, {Juaristi Campillo, J.}, {Julbe, F.}, {Karbevska, L.}, {Kervella, P.}, {Khanna, S.}, {Kontizas, M.}, {Kordopatis, G.}, {Korn, A. J.}, {K\'osp\'al, \'A}, {Kostrzewa-Rutkowska, Z.}, {Kruszy\'{}nska, K.}, {Kun, M.}, {Laizeau, P.}, {Lambert, S.}, {Lanza, A. F.}, {Lasne, Y.}, {Le Campion, J.-F.}, {Lebreton, Y.}, {Lebzelter, T.}, {Leccia, S.}, {Leclerc, N.}, {Lecoeur-Taibi, I.}, {Liao, S.}, {Licata, E. L.}, {Lindstr\o{}m, H. E. P.}, {Lister, T. A.}, {Livanou, E.}, {Lobel, A.}, {Lorca, A.}, {Loup, C.}, {Madrero Pardo, P.}, {Magdaleno Romeo, A.}, {Managau, S.}, {Mann, R. G.}, {Manteiga, M.}, {Marchant, J. M.},
  {Marconi, M.}, {Marcos, J.}, {Marcos Santos, M. M. S.}, {Mar\'{\i}n Pina, D.}, {Marinoni, S.}, {Marocco, F.}, {Marshall, D. J.}, {Martin Polo, L.}, {Mart\'{\i}n-Fleitas, J. M.}, {Marton, G.}, {Mary, N.}, {Masip, A.}, {Massari, D.}, {Mastrobuono-Battisti, A.}, {Mazeh, T.}, {McMillan, P. J.}, {Messina, S.}, {Michalik, D.}, {Millar, N. R.}, {Mints, A.}, {Molina, D.}, {Molinaro, R.}, {Moln\'ar, L.}, {Monari, G.}, {Mongui\'o, M.}, {Montegriffo, P.}, {Montero, A.}, {Mor, R.}, {Mora, A.}, {Morbidelli, R.}, {Morel, T.}, {Morris, D.}, {Muraveva, T.}, {Murphy, C. P.}, {Musella, I.}, {Nagy, Z.}, {Noval, L.}, {Oca\~na, F.}, {Ogden, A.}, {Ordenovic, C.}, {Osinde, J. O.}, {Pagani, C.}, {Pagano, I.}, {Palaversa, L.}, {Palicio, P. A.}, {Pallas-Quintela, L.}, {Panahi, A.}, {Payne-Wardenaar, S.}, {Pe\~nalosa Esteller, X.}, {Penttil\"a, A.}, {Pichon, B.}, {Piersimoni, A. M.}, {Pineau, F.-X.}, {Plachy, E.}, {Plum, G.}, {Poggio, E.}, {Prsa, A.}, {Pulone, L.}, {Racero, E.}, {Ragaini, S.}, {Rainer, M.}, {Raiteri, C. M.},
  {Rambaux, N.}, {Ramos, P.}, {Ramos-Lerate, M.}, {Re Fiorentin, P.}, {Regibo, S.}, {Richards, P. J.}, {Rios Diaz, C.}, {Ripepi, V.}, {Riva, A.}, {Rix, H.-W.}, {Rixon, G.}, {Robichon, N.}, {Robin, A. C.}, {Robin, C.}, {Roelens, M.}, {Rogues, H. R. O.}, {Rohrbasser, L.}, {Romero-G\'omez, M.}, {Rowell, N.}, {Royer, F.}, {Ruz Mieres, D.}, {Rybicki, K. A.}, {Sadowski, G.}, {S\'aez N\'u\~nez, A.}, {Sagrist\`a Sell\'es, A.}, {Sahlmann, J.}, {Salguero, E.}, {Samaras, N.}, {Sanchez Gimenez, V.}, {Sanna, N.}, {Santove\~na, R.}, {Sarasso, M.}, {Schultheis, M.}, {Sciacca, E.}, {Segol, M.}, {Segovia, J. C.}, {S\'egransan, D.}, {Semeux, D.}, {Shahaf, S.}, {Siddiqui, H. I.}, {Siebert, A.}, {Siltala, L.}, {Silvelo, A.}, {Slezak, E.}, {Slezak, I.}, {Smart, R. L.}, {Snaith, O. N.}, {Solano, E.}, {Solitro, F.}, {Souami, D.}, {Souchay, J.}, {Spagna, A.}, {Spina, L.}, {Spoto, F.}, {Steele, I. A.}, {Steidelm\"uller, H.}, {Stephenson, C. A.}, {S\"uveges, M.}, {Surdej, J.}, {Szabados, L.}, {Szegedi-Elek, E.}, {Taris, F.}, {Taylor,
  M. B.}, {Teixeira, R.}, {Tolomei, L.}, {Tonello, N.}, {Torra, F.}, {Torra, J.}, {Torralba Elipe, G.}, {Trabucchi, M.}, {Tsounis, A. T.}, {Turon, C.}, {Ulla, A.}, {Unger, N.}, {Vaillant, M. V.}, {van Dillen, E.}, {van Reeven, W.}, {Vanel, O.}, {Vecchiato, A.}, {Viala, Y.}, {Vicente, D.}, {Voutsinas, S.}, {Weiler, M.}, {Wevers, T.}, {Wyrzykowski, L.}, {Yoldas, A.}, {Yvard, P.}, {Zhao, H.}, {Zorec, J.}, {Zucker, S.}, \& {Zwitter, T.}}]{Gaia2023}
{Gaia Collaboration}, {Vallenari, A.}, {Brown, A. G. A.}, {et~al.} 2023, A\&A, 674, A1, \dodoi{10.1051/0004-6361/202243940}

\bibitem[{Harris {et~al.}(2020)Harris, Millman, van~der Walt, Gommers, Virtanen, Cournapeau, Wieser, Taylor, Berg, Smith, Kern, Picus, Hoyer, van Kerkwijk, Brett, Haldane, del R{\'{i}}o, Wiebe, Peterson, G{\'{e}}rard-Marchant, Sheppard, Reddy, Weckesser, Abbasi, Gohlke, \& Oliphant}]{NumPy}
Harris, C.~R., Millman, K.~J., van~der Walt, S.~J., {et~al.} 2020, Nature, 585, 357, \dodoi{10.1038/s41586-020-2649-2}

\bibitem[{Hunter(2007)}]{Matplotlib}
Hunter, J.~D. 2007, Computing in Science \& Engineering, 9, 90, \dodoi{10.1109/MCSE.2007.55}

\bibitem[{{Karata{\c{s}}} {et~al.}(2023){Karata{\c{s}}}, {{\c{C}}akmak}, {Akkaya Oralhan}, {Bonatto}, {Michel}, \& {Netopil}}]{Karatas2023}
{Karata{\c{s}}}, Y., {{\c{C}}akmak}, H., {Akkaya Oralhan}, {\.I}., {et~al.} 2023, \mnras, 521, 2408, \dodoi{10.1093/mnras/stad565}

\bibitem[{{Knigge} {et~al.}(2009){Knigge}, {Leigh}, \& {Sills}}]{Knigge2009}
{Knigge}, C., {Leigh}, N., \& {Sills}, A. 2009, \nat, 457, 288, \dodoi{10.1038/nature07635}

\bibitem[{{Li} {et~al.}(2020){Li}, {Shao}, {Li}, {Yu}, {Zhong}, \& {Chen}}]{LiBinaries2020}
{Li}, L., {Shao}, Z., {Li}, Z.-Z., {et~al.} 2020, \apj, 901, 49, \dodoi{10.3847/1538-4357/abaef3}

\bibitem[{{Mathieu}(2000)}]{WOCS2000}
{Mathieu}, R.~D. 2000, in Astronomical Society of the Pacific Conference Series, Vol. 198, Stellar Clusters and Associations: Convection, Rotation, and Dynamos, ed. R.~{Pallavicini}, G.~{Micela}, \& S.~{Sciortino}, 517

\bibitem[{McLachlan \& Peel(2000)}]{McLachlanPeel2000}
McLachlan, G.~J., \& Peel, D. 2000, Probability and Statistics -- Applied Probability and Statistics Section, Vol. 299, Finite mixture models (New York: Wiley)

\bibitem[{{Meylan} \& {Heggie}(1997)}]{MeylanHeggie1997}
{Meylan}, G., \& {Heggie}, D.~C. 1997, \aapr, 8, 1, \dodoi{10.1007/s001590050008}

\bibitem[{{Milliman} {et~al.}(2014){Milliman}, {Mathieu}, {Geller}, {Gosnell}, {Meibom}, \& {Platais}}]{Milliman2014}
{Milliman}, K.~E., {Mathieu}, R.~D., {Geller}, A.~M., {et~al.} 2014, {VizieR Online Data Catalog: WIYN open cluster study. LX. RV survey of NGC 6819 (Milliman+, 2014)}, VizieR On-line Data Catalog: J/AJ/148/38. Originally published in: 2014AJ....148...38M, \dodoi{10.26093/cds/vizier.51480038}

\bibitem[{{Nine} {et~al.}(2020){Nine}, {Milliman}, {Mathieu}, {Geller}, {Leiner}, {Platais}, \& {Tofflemire}}]{Nine2020}
{Nine}, A.~C., {Milliman}, K.~E., {Mathieu}, R.~D., {et~al.} 2020, \aj, 160, 169, \dodoi{10.3847/1538-3881/abad3b}

\bibitem[{Panthi {et~al.}(2023)Panthi, Vaidya, Vernekar, Subramaniam, Jadhav, \& Agarwal}]{Panthi2023}
Panthi, A., Vaidya, K., Vernekar, N., {et~al.} 2023, Monthly Notices of the Royal Astronomical Society, 527, 8325, \dodoi{10.1093/mnras/stad3750}

\bibitem[{{Rain} {et~al.}(2021){Rain}, {Ahumada}, \& {Carraro}}]{Rain2021}
{Rain}, M.~J., {Ahumada}, J.~A., \& {Carraro}, G. 2021, \aap, 650, A67, \dodoi{10.1051/0004-6361/202040072}

\bibitem[{{Rain} {et~al.}(2020){Rain}, {Carraro}, {Ahumada}, {Villanova}, {Boffin}, {Monaco}, \& {Beccari}}]{Rain2020}
{Rain}, M.~J., {Carraro}, G., {Ahumada}, J.~A., {et~al.} 2020, \aj, 159, 59, \dodoi{10.3847/1538-3881/ab5f0b}

\bibitem[{{Rao} {et~al.}(2023){Rao}, {Bhattacharya}, {Vaidya}, \& {Agarwal}}]{RaoBerkeley172023}
{Rao}, K.~K., {Bhattacharya}, S., {Vaidya}, K., \& {Agarwal}, M. 2023, \mnras, 518, L7, \dodoi{10.1093/mnrasl/slac122}

\bibitem[{Rao {et~al.}(2023)Rao, Vaidya, Agarwal, Balan, \& Bhattacharya}]{Rao2023}
Rao, K.~K., Vaidya, K., Agarwal, M., Balan, S., \& Bhattacharya, S. 2023, Monthly Notices of the Royal Astronomical Society, 526, 1057, \dodoi{10.1093/mnras/stad2755}

\bibitem[{{Rao} {et~al.}(2021){Rao}, {Vaidya}, {Agarwal}, \& {Bhattacharya}}]{Rao2021}
{Rao}, K.~K., {Vaidya}, K., {Agarwal}, M., \& {Bhattacharya}, S. 2021, \mnras, 508, 4919, \dodoi{10.1093/mnras/stab2894}

\bibitem[{{Sanna} {et~al.}(2014){Sanna}, {Dalessandro}, {Ferraro}, {Lanzoni}, {Miocchi}, \& {O'Connell}}]{Sanna2014}
{Sanna}, N., {Dalessandro}, E., {Ferraro}, F.~R., {et~al.} 2014, \apj, 780, 90, \dodoi{10.1088/0004-637X/780/1/90}

\bibitem[{Scholz \& Stephens(1987)}]{ADTest}
Scholz, F.~W., \& Stephens, M.~A. 1987, Journal of the American Statistical Association, 82, 918.
\newblock \url{http://www.jstor.org/stable/2288805}

\bibitem[{{Shara} {et~al.}(1997){Shara}, {Saffer}, \& {Livio}}]{Shara1997}
{Shara}, M.~M., {Saffer}, R.~A., \& {Livio}, M. 1997, \apjl, 489, L59, \dodoi{10.1086/310952}

\bibitem[{{Stryker}(1993)}]{Stryker1993}
{Stryker}, L.~L. 1993, \pasp, 105, 1081, \dodoi{10.1086/133286}

\bibitem[{{Vaidya} {et~al.}(2022){Vaidya}, {Panthi}, {Agarwal}, {Pandey}, {Rao}, {Jadhav}, \& {Subramaniam}}]{Vaidya2022}
{Vaidya}, K., {Panthi}, A., {Agarwal}, M., {et~al.} 2022, \mnras, 511, 2274, \dodoi{10.1093/mnras/stac207}

\bibitem[{{Vaidya} {et~al.}(2020){Vaidya}, {Rao}, {Agarwal}, \& {Bhattacharya}}]{Vaidya2020}
{Vaidya}, K., {Rao}, K.~K., {Agarwal}, M., \& {Bhattacharya}, S. 2020, \mnras, 496, 2402, \dodoi{10.1093/mnras/staa1667}

\bibitem[{{van Groeningen} {et~al.}(2023){van Groeningen}, {Castro-Ginard}, {Brown}, {Casamiquela}, \& {Jordi}}]{VG2023}
{van Groeningen}, M.~G.~J., {Castro-Ginard}, A., {Brown}, A.~G.~A., {Casamiquela}, L., \& {Jordi}, C. 2023, \aap, 675, A68, \dodoi{10.1051/0004-6361/202345952}

\bibitem[{Virtanen {et~al.}(2020)Virtanen, Gommers, Oliphant, Haberland, Reddy, Cournapeau, Burovski, Peterson, Weckesser, Bright, {van der Walt}, Brett, Wilson, Millman, Mayorov, Nelson, Jones, Kern, Larson, Carey, Polat, Feng, Moore, {VanderPlas}, Laxalde, Perktold, Cimrman, Henriksen, Quintero, Harris, Archibald, Ribeiro, Pedregosa, {van Mulbregt}, \& {SciPy 1.0 Contributors}}]{SciPy}
Virtanen, P., Gommers, R., Oliphant, T.~E., {et~al.} 2020, Nature Methods, 17, 261, \dodoi{10.1038/s41592-019-0686-2}

\bibitem[{{Wu} {et~al.}(2007){Wu}, {Zhou}, {Ma}, {Jiang}, {Chen}, \& {Wu}}]{WuNGC77892007}
{Wu}, Z.-Y., {Zhou}, X., {Ma}, J., {et~al.} 2007, \aj, 133, 2061, \dodoi{10.1086/512189}

\bibitem[{{Zwicker} {et~al.}(2023){Zwicker}, {Geller}, {Childs}, {Motherway}, \& {von Hippel}}]{Zwicker2023}
{Zwicker}, C., {Geller}, A.~M., {Childs}, A.~C., {Motherway}, E., \& {von Hippel}, T. 2023, arXiv e-prints, arXiv:2308.15582, \dodoi{10.48550/arXiv.2308.15582}

\end{thebibliography}
\bibliographystyle{aasjournal}

\renewcommand{\thefigure}{A\arabic{figure}}

\setcounter{figure}{0}

\begin{figure*}
    \centering
    \includegraphics[width=\linewidth]{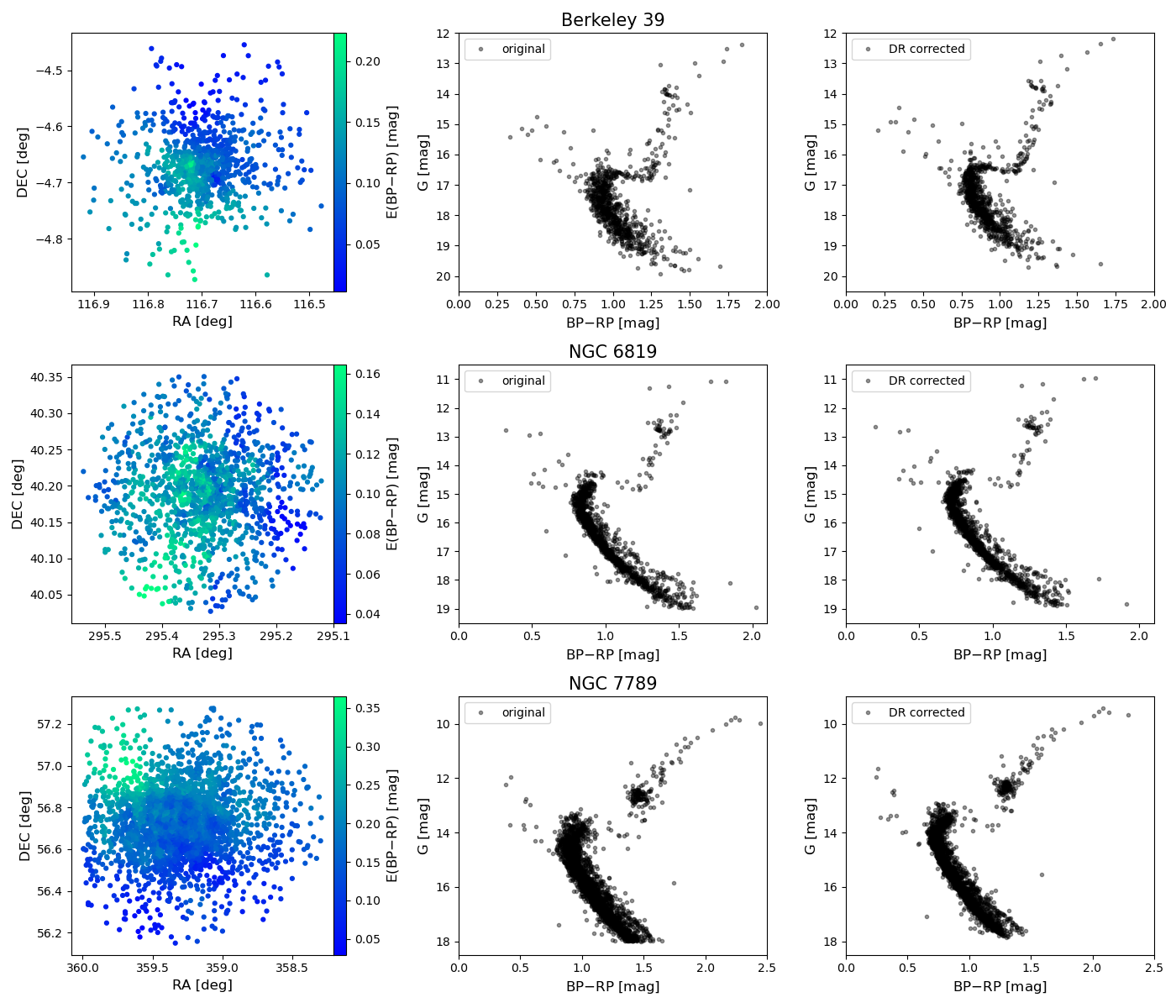}
    \caption{Differential Reddening correction for Berkeley 39, NGC 6819 and NGC 7789 (from top to bottom) with the reddening map, observed CMD and the reddening corrected CMD for each cluster (from left to right).}
    \label{fig:dr_others}
\end{figure*}


\end{document}